\documentclass{aa}
\usepackage{natbib}
\bibpunct{(}{)}{;}{a}{}{,} % to follow the A&A style

\usepackage{graphicx}

\usepackage{amssymb,amsmath,bm,braket}
\usepackage{delarray}
\usepackage{mathrsfs}
\usepackage{xcolor}
\usepackage{soul}
\usepackage[varg]{txfonts}
\usepackage{hyperref}
\usepackage{multirow}
\hypersetup{
  colorlinks   = true, %Colours links instead of ugly boxes
  urlcolor     = blue, %Colour for external hyperlinks
  linkcolor    = blue, %Colour of internal links
  citecolor    = blue  %Colour of citations
}

\newcommand{\rr}{{\bm r}}
\setcitestyle{notesep={ }}

\begin{document} 

\title{A numerical approach for modelling the polarisation signals\\ of strong resonance lines with partial frequency redistribution}
\subtitle{Numerical applications to two-term atoms and plane-parallel atmospheres}

\authorrunning{Riva et al.}
\titlerunning{Polarisation of resonance lines with PRD}

\author{Fabio Riva\inst{1,2}
       \and
       Gioele Janett\inst{1,2}
       \and
       Luca Belluzzi\inst{1,2,3}
       \and
       Tanaus\'u del Pino Alem\'an\inst{4,5}
       \and
       Ernest Alsina Ballester\inst{4,5}
       \and\\
       Javier Trujillo Bueno\inst{4,5,6}
       \and
       Pietro Benedusi\inst{2}
       \and
       Simone Riva\inst{2}
       \and
       Rolf Krause\inst{2,7,8}
       }

\institute{
        Istituto ricerche solari Aldo e Cele Daccò (IRSOL), Faculty of Informatics, Università della Svizzera italiana 
        (USI), CH-6605 Locarno, Switzerland 
        \and
        Euler Institute, Universit\`a della Svizzera italiana (USI), CH-6900 Lugano, Switzerland
        \and
        Institut f\"ur Sonnenphysik (KIS), D-79104 Freiburg i.~Br., Germany
        \and
        Instituto de Astrof\'isica de Canarias, E-38205 La Laguna, Tenerife, Spain
        \and
        Departamento de Astrofísica, Universidad de La Laguna, E-38206 La Laguna, Tenerife, Spain
        \and
        Consejo Superior de Investigaciones Cient\'ificas, Spain
        \and
        FernUni, CH-3900 Brig, Switzerland
        \and{AMCS, KAUST, King Abdullah University of Science and Technology, Thuwal, Saudi-Arabia}\\
        \email{fabio.riva@irsol.usi.ch}
        }

\date{Received xxx; accepted yyy}
 
\abstract
{}
{
The main goal of this paper is to present an accurate and efficient numerical strategy for solving the radiative transfer problem for polarised radiation in strong resonance lines forming out of local thermodynamic equilibrium, taking angle-dependent~(AD) partial frequency redistribution~(PRD) effects and $J$-state interference into account.
We consider the polarisation produced both by the Zeeman effect and by the scattering of anisotropic radiation, 
along with its sensitivity to the Hanle and magneto-optical effects.
}
{
We introduce a formalism that allows treating both a two-level and a two-term atom in the presence of arbitrary magnetic and bulk velocity fields. The problem is formulated by treating the population of the lower level/term as a fixed input parameter. This approach makes the problem linear with respect to the radiation field, enabling the application of efficient matrix-free preconditioned iterative methods for its solution.
Additionally, the computation of the scattering emissivity in the comoving frame, together with a careful choice of the angular and spectral quadrature nodes, 
allow us to speed up the calculations by reducing the number of evaluations of the redistribution functions.
}
{
The proposed solution strategy is
applied to synthesise the Stokes profiles of the Mg~{\sc ii} h\&k doublet and the H~{\sc i} Ly-$\alpha$ line in 1D semi-empirical models.
The results demonstrate that the method is both fast and accurate.
A comparison with calculations from HanleRT-TIC
displays an overall good agreement, thereby
validating our solution strategy.
Moreover, for the wavelength-integrated polarisation profiles of the H~{\sc i} Ly-$\alpha$ line, we find an excellent agreement between the results obtained including PRD effects in their general AD description and those obtained considering the angle-averaged simplifying approximation.
}
{}

\keywords{Radiative transfer -- Scattering -- polarisation -- Sun: atmosphere -- Methods: numerical}

\maketitle

\section{Introduction}\label{sec:introduction}
The magnetic field is known to play a key role in the solar chromosphere and chromosphere-corona transition region, but its inference in these layers is notoriously challenging.
This difficulty is because the information about these atmospheric layers is mainly contained in strong ultraviolet (UV) lines
and the sensitivity of the Zeeman effect diminishes at short wavelengths and in hot plasmas.
One of the most promising strategies to overcome this issue is to exploit, in addition to the Zeeman effect, the magnetic sensitivity of the linear polarisation signals that the scattering of anisotropic radiation produces in strong resonance lines of the solar spectrum \citep[e.g.][]{trujillo_bueno2022}.
Indeed, conspicuous scattering polarisation signals are observed in several strong chromospheric lines, both in the visible 
\citep[e.g.][]{Gandorfer00,Gandorfer02,Gandorfer05} and in the UV \citep[e.g.][]{kano2017,rachmeler2022}. 
These signals, which usually extend into the line wings, are sensitive to the magnetic field through the joint action of two physical mechanisms: the Hanle effect \citep[e.g.][]{landi_deglinnocenti+landolfi2004} and magneto-optical (MO) effects \citep[e.g.][]{alsinaballester2016,delpinoaleman2016}.
The former operates in the core of spectral lines, the latter in the wings.
With respect to the Zeeman effect, the Hanle effect allows for the detection of significantly weaker magnetic fields (typically in the range $1-100$\,G). 
Moreover, its sensitivity does not decrease for wavelengths towards the blue or when observing very hot plasmas \citep[e.g.][]{trujillo_bueno2014}.
Besides, MO effects 
are effective in modifying the scattering polarisation wings of strong resonance lines in a range of magnetic field strengths similar to that of the Hanle effect~\citep[][]{alsinaballester2016}.
Notably, various instruments have been recently developed, or proposed, to exploit the diagnostic potential of scattering polarisation in strong UV lines, such as the series of CLASP sounding rocket experiments,~\citep{kobayashi2012,McKenzie2019,mckenzie21} and the proposed Chromospheric Magnetism Explorer~(CMEx) satellite mission.\footnote{\href{https://www.nasa.gov/news-release/nasa-selects-four-small-explorer-mission-concept-studies}{https://www.nasa.gov/news-release/nasa-selects-four-small-explorer-mission-concept-studies}}

To accurately model the scattering polarisation signals of strong resonance lines, as measured by such instruments,
it is necessary to solve the radiative transfer (RT) problem for polarised radiation out of local thermodynamic equilibrium (NLTE conditions), taking partial frequency redistribution (PRD) effects in scattering processes into account.
Additionally, in the case of atomic multiplets, another important element to consider is quantum interference between different fine-structure levels of the same term, generally referred to as $J$-state interference \citep[e.g.][]{belluzzi11,belluzzi2012a,belluzzi2012b}.
Various theoretical approaches 
for describing the generation and transfer of polarised radiation accounting for PRD effects and $J$-state interference have been proposed during the last decades 
\citep[see, e.g. Sect.~1 of][and references therein]{casini2014}.
Here, we mention the one by~\citet{casini2014,casini2017a,casini2017b} and the one by \citet{bommier2017}, which converge to the same expressions when considering a two-term atom (i.e. an atomic model composed of two terms and including $J$-state interference) with an unpolarised lower term~\citep{casini2017b}.
The former is presently implemented in the HanleRT-TIC code~\citep{delpinoaleman2016,delpinoaleman2020,li22}.\footnote{We note that HanleRT-TIC allows considering both two-term and multi-term lambda type~\citep{Casini_2016,Casini_2016b} atomic models.}
The latter is implemented in the code described in \citet{alsinaballester2022}
and is also used in this work. 

Modelling PRD effects is notoriously demanding from a computational point of view, even when just considering a 1D plane-parallel atmospheric geometry or neglecting polarisation.
To simplify the problem, the so-called angle-averaged (AA) approximation \citep[e.g.][]{mihalas1978,rees1982,bommier1997b} has been widely applied in the past.
The first results of NLTE RT calculations for scattering polarisation, in 1D plane-parallel models of the solar atmosphere, including PRD effects in their most general angle-dependent~(AD) formulation, have been obtained only very recently \citep[e.g.][]{delpinoaleman2020,janett2021b,riva2023,guerreiro24}.
This is presently considered the state of the art in the numerical modelling of Hanle-Zeeman signals, and it is still regarded as an extremely complex and challenging computational problem. 
Consequently, it is crucial to devise numerical methods tailored for the considered problem, which are at the same time fast and accurate.
This will help us better understand the relative impact of the various physical processes at play on the observed polarisation signals and develop reliable inversion techniques targeted at extracting information on the upper layers of the solar atmosphere.

In this paper, we present the solution strategy that we have developed to solve the NLTE RT problem for polarised radiation taking AD PRD effects into account. Our approach allows considering both two-level and two-term atomic models, with an unpolarised and infinitely-sharp lower level/term, in the presence of arbitrary magnetic and bulk velocity fields.
As the computational bottleneck of the problem is the evaluation of the contribution of scattering processes to the line emissivity and, in particular, the large number of evaluations of the so-called redistribution functions, we devise a strategy to minimise this number and speed up the calculations.
This strategy is implemented in a code 
named TRAnsfer of Polarized radiation in Plane-Parallel models with PRD (TRAP$^4$), which allows solving the problem considering 1D plane-parallel atmospheric models. The TRAP$^4$ code is then applied to synthesise the Stokes profiles of the Mg~{\sc ii} h\&k doublet and the H~{\sc i} Ly-$\alpha$ line.

The paper is structured as follows. 
In Sect.~\ref{sec:problem}, we present the main equations inherent to the NLTE RT problem for polarised radiation, along with the redistribution matrix formalism for describing PRD effects. 
We also discuss our main assumptions and approximations, and we present the discretisation and algebraic formulation of the problem used in this work.
In Sect.~\ref{sec:formal_solution}, we 
discuss the formal solution of the RT equation.
Then, in Sect.~\ref{sec:calculation_eps}, we describe the numerical calculation of the emissivity, focusing on the quadrature of the so-called scattering integral. 
Section~\ref{sec:iterative} exposes the applied matrix-free iterative solution strategy. 
In Sect.~\ref{sec:results}, we report on the verification of our calculations, the validation of our solution strategy, and the synthesis of wavelength-integrated linear polarisation profiles of the H~{\sc i} Ly-$\alpha$ line. 
Finally, we provide remarks and conclusions in Sect.~\ref{sec:conclusions}.

\section{Formulation of the problem}\label{sec:problem}
In this section, we first present the most relevant equations describing the considered NLTE RT problem for polarised radiation.
Second, we expose the main assumption of the proposed approach, under which the problem is linear with respect to the radiation field.
Third, we present a suitable discretisation of the linear problem.
Finally, we outline the ensuing algebraic formulation, which proves to be particularly simple and accessible.

\subsection{NLTE RT problem}

The intensity and polarisation of a beam of radiation are fully described by the four-component Stokes vector
\begin{equation*}\label{eq:Stokes}
        \bm{I} = (I,Q,U,V)^T = (I_1, I_2, I_3, I_4)^T \in \mathbb R_+\times \mathbb R^3,
\end{equation*}
where $I$ is the specific intensity, while $Q$, $U$, and $V$ encode the polarisation, always satisfying the condition $I \ge \sqrt{Q^2 + U^2 + V^2}$.
For a given medium, such as the plasma of a stellar atmosphere, 
the variation of a radiation beam of frequency~$\nu\in\mathbb R_+$ and direction~$\vec{\Omega}$ 
at a spatial location~$\rr\in D$, with $D\subset \mathbb R^3$, 
is described by the following RT equation
\begin{equation}\label{eq:RT}
        \vec{\nabla}_{{\vec{\Omega}}}\bm{I}(\rr,\vec{\Omega},\nu) = 
        -K(\rr,\vec{\Omega},\nu) 
        \bm{I}(\rr,\vec{\Omega},\nu) + 
        \pmb{\varepsilon}(\rr,\vec{\Omega},\nu),
\end{equation}
where $\vec{\nabla}_{\vec{\Omega}}=\vec{\Omega}\cdot\partial/\partial\bf{\rr}$ denotes the directional derivative along the unit vector $\vec{\Omega}$.
The propagation matrix $K\in\mathbb R^{4\times4}$
quantifies how the medium attenuates the intensity for all polarisation states (absorption), and
how it differentially absorbs (dichroism) and couples (dispersion) the different Stokes parameters.
The emission vector
\begin{equation*}
        \pmb{\varepsilon} = 
        (\varepsilon_1, \varepsilon_2, \varepsilon_3, \varepsilon_4)^T \in \mathbb R_+\times \mathbb R^3
\end{equation*}
represents the radiation emitted by the plasma in the four 
Stokes parameters.
The RT coefficients $K$ and $\pmb{\varepsilon}$ include all the relevant contributions of both line and continuum processes. Hereafter, we only present the former, as this is the central topic of this paper, while we refer to Appendix~\ref{app:continuum} for a discussion of the latter.

The contribution of line processes depends 
on the state of the atom giving rise to the considered spectral line.
This needs to be determined by solving the statistical equilibrium (SE) equations, which describe the interaction of the atom with the radiation field (radiative processes), other particles (collisional processes), and the possible presence of external electric and/or magnetic fields.
The NLTE RT problem thus consists in finding a self-consistent solution of the RT equation, Eq.~\eqref{eq:RT}, for the radiation field $\bm{I}$ and the SE equations for the atomic system.
This problem is in general integro-differential, non-local, and, because of the dependence of the radiative rates in the SE equations on the radiation field $\bm{I}$, 
non-linear.

\subsection{Atomic models}\label{sec:atomic_model}
In this work, we consider an atomic model composed of two terms \citep[two-term atom, see Sect.~7.6 of][]{landi_deglinnocenti+landolfi2004}, which includes the two-level atom as a particular case.
The formalism that we introduce in the following allows treating both atomic models.

Assuming $L-S$ coupling, each term of a two-term atom
is specified by a set of inner quantum numbers $\beta$, the quantum number for the orbital angular momentum $L$, and the quantum number for the electronic spin $S$.
In the absence of magnetic fields, the energy eigenstates are given by $\ket{\beta LS JM}$, where $J$ is the total angular momentum quantum number and $M\in\{-J,-J+1,...,J\}$ is the magnetic quantum number.
The presence of an external magnetic field induces a splitting between the magnetic sublevels.
If the magnetic splitting is much smaller than the separation between the different $J$-levels, it can be calculated in the Zeeman effect regime, and it is thus linear with respect to the magnetic field strength $B$.
On the other hand, if the magnetic splitting is comparable to the separation between different $J$-levels, one has to consider the Paschen-Back effect. 
In this case, the splitting is not linear with $B$ and $J$ ceases to be a good quantum number \citep[e.g.][]{landi_deglinnocenti+landolfi2004}.
In the present work, we account for the Paschen-Back effect and we introduce a label $j$ to distinguish between different states with the same quantum numbers $\beta$, $L$, $S$, and $M$.
We recall that a two-term atomic model accounts for $J$-state interference within each term.
The case of a two-level atom can be easily recovered by setting $S=0$, so that each term is composed of a single $J$-level, with $J=L$.

Hereafter, we assume that the lower level/term of the considered atomic system is infinitely sharp.
This is a very good assumption for resonance lines, as their lower level (ground or metastable) has a very long lifetime. 
Moreover, we neglect atomic polarisation in the lower level/term.
This assumption, which is always correct for levels with $J=0$ and $J=1/2$,\footnote{Levels with $J=1/2$ cannot carry atomic alignment by definition, but can carry atomic orientation if the incident radiation has net circular polarisation.} is generally good for any long-lived level in a sufficiently dense plasma, due to the depolarising effect of elastic collisions.
The assumptions of unpolarised and infinitely sharp lower level/term are thus very good for strong resonance lines such as Ca~{\sc i} 4227 and Sr~{\sc ii} 4077, 
whose polarisation signals are suitably modelled considering a two-level atom, as well as Mg~{\sc ii} h\&k, H~{\sc i} Ly-$\alpha$, and He~{\sc ii} 304, which
require a two-term atomic model.

The explicit expressions of the line contributions to the elements of the propagation matrix $K$ for a two-level and a two-term atom with unpolarised lower level/term can be found in Chapter~7 of \citet{landi_deglinnocenti+landolfi2004}.
The line contribution to the emission vector for the same atomic models is discussed in the next section.
Noticing that a two-term atom is formally equivalent to a two-level atom with hyperfine structure \citep[e.g.][]{landi_deglinnocenti+landolfi2004},
the framework described in this paper can be applied to both atomic models, as detailed in \citet[][]{janett2023}.

\subsection{Line emissivity}
\label{sec:scattering}

Neglecting stimulated emission, the SE equations for two-level or two-term atomic models with an unpolarised lower level/term have an analytic solution, and it is thus possible to apply the redistribution matrix formalism to describe PRD effects \citep[e.g.][]{mihalas1978}.
In this case, the line contribution to the emission vector can be written as the sum of two terms describing the contributions from atoms that are either radiatively
excited (scattering term, label `sc') or collisionally excited (thermal term, label `th'), that is,
\begin{equation*}
	\pmb{\varepsilon}^\ell(\rr,\vec{\Omega},\nu) = 
	\pmb{\varepsilon}^{\ell,\text{sc}}(\rr,\vec{\Omega},\nu) + 
	\pmb{\varepsilon}^{\ell,\text{th}}(\rr,\vec{\Omega},\nu),
\end{equation*}
where the superscript $\ell$ denotes line processes.

Under the assumption of isotropic inelastic collisions, the thermal term is given by 
$\pmb{\varepsilon}^{\ell,\text{th}}(\rr,\vec{\Omega},\nu)=\epsilon(\rr)W_\mathrm{T}(\rr)\vec{\eta}^\ell(\rr,\vec{\Omega},\nu)$, where 
$\epsilon(\rr) = \Gamma_{\rm I}(\rr)/\left[\Gamma_{\rm R} + \Gamma_{\rm I}(\rr)\right]$
is the photon destruction probability,
being $\Gamma_{\rm R}$ and $\Gamma_{\rm I}$ the line broadening constants due respectively to radiative decays and inelastic collisions, $W_\mathrm{T}$ is the Planck function in the Wien limit (consistently with the assumption of neglecting stimulated emission),\footnote{For simplicity, we neglect the dependence of the Wien function on frequency in the spectral interval of the line.}
and the vector $\vec{\eta}^\ell$ gathers the absorption and dichroism coefficients~\citep{landi_deglinnocenti+landolfi2004}.
Moreover, using the redistribution matrix formalism, the scattering term is directly related to the radiation field that illuminates the atom through the scattering integral
\begin{equation}\label{eq:eps_sc_l}
        \pmb{\varepsilon}^{{\ell,\rm sc}}(\rr,\vec{\Omega},\nu) \!=\! 
        k(\rr)\!\! \int\!\! {\rm d} \nu'\! 
        \oint\! \frac{{\rm d} \vec{\Omega}'}{4 \pi}
        R(\rr,\vec{\Omega}',\vec{\Omega},\nu',\nu) 
        \bm{I}(\rr,\vec{\Omega}',\nu'),
\end{equation}
where $k$ is the frequency-integrated absorption coefficient 
and $R\in\mathbb R^{4\times4}$ the redistribution matrix, which encodes an analytic solution of the SE equations. 
Here, primed and unprimed quantities refer to the incident and scattered radiation, respectively.

The most general form of the redistribution matrix for an atomic system with infinitely 
sharp lower states is expressed by the linear combination of two terms \citep[e.g.][]{bommier1997a,bommier1997b,bommier2017}
\begin{equation*}
R(\rr,\vec{\Omega}',\vec{\Omega},\nu',\nu) = R^{\scriptscriptstyle \rm II}(\rr,\vec{\Omega}',\vec{\Omega},\nu',\nu) +
R^{\scriptscriptstyle \rm III}(\rr,\vec{\Omega}',\vec{\Omega},\nu',\nu),
\end{equation*}
where $R^{\scriptscriptstyle \rm II}$ and $R^{\scriptscriptstyle \rm III}$ describe scattering processes 
that are coherent in frequency and totally uncorrelated, respectively, in the atomic rest frame.
Using the formalism of the irreducible spherical tensors for polarimetry
\citep[see Chapter~5 of][]{landi_deglinnocenti+landolfi2004}, each redistribution matrix can be written as
\begin{equation*}
	R^{\scriptscriptstyle \rm X}(\rr,\vec{\Omega}',
	\vec{\Omega},\nu',\nu) = \!\! 
	\sum_{K,K',Q}\!\!\mathcal{R}^{{\scriptscriptstyle \rm X},KK'}_Q(\rr,\vec{\Omega}', 
	\vec{\Omega},\nu',\nu) 
	\mathcal{P}^{KK'}_{Q}(\rr,\vec{\Omega}',\vec{\Omega}),
\end{equation*}
with ${\rm X = II,III}$, $\mathcal{R}^{{\scriptscriptstyle \rm X},KK'}_Q \in \mathbb{C}$ the redistribution function, and
$\mathcal{P}^{KK'}_{Q} \in \mathbb{C}^{4\times 4}$ the scattering phase matrix.
The explicit expression of the scattering phase matrix can be found in~\citet{alsinaballester2022}.

\subsection{Redistribution functions}
\label{sec:redistribution_functions}

Hereafter, we outline the expressions of the redistribution functions considered in this work, as this is a crucial element of our optimisation strategy. 
To treat both a two-level and a two-term atomic model within the same formalism, we introduce 
the indexes $k_u$ and $k_\ell$ to characterise the upper (label `$u$') and lower (label `$\ell$') states of the considered atom. Specifically, we have $k_u=M_u$ and $k_\ell=M_\ell$ for a two-level atom, and $k_u=j_uM_u$ and $k_\ell=j_\ell M_\ell$ for a two-term atom.\footnote{Although not considered in this work, the present formalism can be simply extended to a two-term atom with hyperfine structure
by taking $k_u$ and $k_\ell$ as the quantum numbers identifying the upper and lower states, respectively,
of such an atomic model~\citep[see, e.g.][]{alsinaballester2022}.}
Moreover, since the computation of $\mathcal{R}^{{\scriptscriptstyle \rm X},KK'}_Q$ is local in space, in this subsection we neglect the dependence of any quantity on $\rr$ for ease of notation.

The redistribution function $\mathcal{R}^{{\scriptscriptstyle \rm II},KK'}_Q$ in its general AD description and in the observer's frame can be written as
\begin{align}
\mathcal R_Q^{{\scriptscriptstyle \rm II},KK'}(\vec{\Omega}',\vec{\Omega},\nu',\nu)=&\sum_{k_uk_u'}\sum_{k_\ell k_\ell'}A_Q^{KK'}(k_u,k_u',k_\ell,k_\ell')\nonumber\\
&\times\left[G(\vec{\Omega}',\vec{\Omega},\nu',\nu,k_u',k_\ell,k_\ell')\right.\nonumber\\
&\left.+\,\,\, G(\vec{\Omega}',\vec{\Omega},\nu',\nu,k_u,k_\ell,k_\ell')^*\right],\label{eq:RQIIKKp}
\end{align}
where the function $G$ is defined as
\begin{align}
G(\vec{\Omega}',&\vec{\Omega},\nu',\nu,k_u,k_\ell,k_\ell')=\nonumber\\
&F\left(\Theta,u(\nu')+u_\mathrm{b}(\vec{\Omega}')+u_{k_u,k_\ell},u(\nu)+u_\mathrm{b}(\vec{\Omega})+u_{k_u,k_\ell'}\right),\label{eq:G}
\end{align}
with
\begin{align*}
F(\Theta,u',u)=\,&\frac{1}{\sin(\Theta)}\exp\left[-\left(\frac{u-u'}{2\sin(\Theta/2)}\right)^2\right]\nonumber\\
&\times W\left(\frac{a}{\cos(\Theta/2)},\frac{u+u'}{2\cos(\Theta/2)}\right),
\end{align*}
while we refer to Appendix~\ref{app:AQKKp} for the definition of $A_Q^{KK'}$.
Here, $\Theta=\arccos(\vec{\Omega} \cdot \vec{\Omega}')$ is the angle between the incoming ${\bf\Omega}'$ and outgoing ${\bf\Omega}$ directions (scattering angle), $W$ is the Faddeeva function, and $a=(\Gamma_\mathrm{R}+\Gamma_\mathrm{I}+\Gamma_\mathrm{E})/(4\pi\Delta\nu_\mathrm{D})$ is the damping constant, 
being $\Gamma_{\rm E}$ the line broadening constant due to elastic collisions.
In Eq.~\eqref{eq:G},
\begin{equation*}
u(\nu)=\frac{\nu_0-\nu}{\Delta\nu_\mathrm{D}},\ \ 
u_\mathrm{b}(\vec{\Omega})=\frac{\nu_0}{c}\frac{\bm{v}_\mathrm{b}\cdot\vec{\Omega}}{\Delta\nu_\mathrm{D}},\ \
u_{k_u,k_\ell}=\frac{\nu_{k_u,k_\ell}-\nu_0}{\Delta\nu_\mathrm{D}},
\end{equation*} 
are the reduced frequency, the reduced Doppler shift frequency due to the local plasma bulk velocity $\bm{v}_\mathrm{b}$,\footnote{In the definition of $u_\mathrm{b}$, we assume that the bulk velocity $\bm{v}_\mathrm{b}$ is non-relativistic. Moreover, we assume that the frequency interval of the line/multiplet (centred in $\nu_0$) is sufficiently small, so that 
$\nu\,\vec{v}_{\rm b} \cdot \vec{\Omega} / c\approx\nu_0 \vec{v}_{\rm b} \cdot \vec{\Omega} / c$.} 
and the reduced frequency for the transition between the upper state $|k_u\rangle$ and the lower state $|k_\ell\rangle$, respectively, being $\nu_0$ a reference frequency for the considered spectral line or multiplet,
$\Delta \nu_D$ the Doppler width in frequency units,
$c$ the speed of light,
and $\nu_{k_u,k_\ell} = [E(k_u) - E(k_\ell)]/h$,
with $E$ and $h$ the energy of the considered state and the Planck constant, respectively.

In the definition of $\mathcal{R}^{{\scriptscriptstyle \rm II},KK'}_Q$, Eq.~\eqref{eq:RQIIKKp}, the function $G$
locally couples all frequencies and directions of incident and scattered radiation, making 
the evaluation of Eq.~\eqref{eq:eps_sc_l} computationally very demanding.
However, we note that, in the absence of bulk velocities (or when evaluated in the comoving frame, see Sect.~\ref{sec:bulkvel}), $G$ does not depend on all the couples $(\vec{\Omega}',\vec{\Omega})$, but only on the scattering angle $\Theta$.
Since evaluating the Faddeeva function is computationally expensive, this property is particularly useful for accelerating the calculation of the emissivity, as detailed in Sect.~\ref{sec:angular_quadrature}.

As far as the $\mathcal R^{\scriptscriptstyle \rm III}$ redistribution function is concerned, we consider its simplified expression under the assumption 
that the scattering processes that it describes are completely uncorrelated not only in the atomic rest frame, but also in the observer's frame. 
This approximation provides accurate results while ensuring a drastic reduction of the computational cost \citep[e.g.][]{sampoorna2017,riva2023}.
The redistribution function $\mathcal R^{\scriptscriptstyle \rm III}$ can then be written in a form similar to Eq.~\eqref{eq:RQIIKKp}:
\begin{align}
\mathcal R_Q^{{\scriptscriptstyle \rm III},KK'}(\vec{\Omega}',\vec{\Omega},\nu',\nu)=&\sum_{k_uk_u'}\sum_{k_\ell k_\ell'}B_Q^{KK'}(k_u,k_u',k_\ell,k_\ell')\nonumber\\
&\times\left[W\left(a,u(\nu')+u_\mathrm{b}(\vec{\Omega}')+u_{k_u',k_\ell}\right)\right.\nonumber\\
&\left. +W\left( a,u(\nu')+u_\mathrm{b}(\vec{\Omega}')+u_{k_u,k_\ell} \right) ^*\right]\nonumber\\
&\times\left[W\left(a,u(\nu)+u_\mathrm{b}(\vec{\Omega})+u_{k_u',k_\ell'}\right)\right.\nonumber\\
&\left.+W\left(a,u(\nu)+u_\mathrm{b}(\vec{\Omega})+u_{k_u,k_\ell'}\right)^*\right],\label{eq:RIIIQKKp}
\end{align}
where the quantity $B_Q^{KK'}$ is defined in Appendix~\ref{app:AQKKp}. 
Notably, in this expression all incident directions and frequencies are decoupled from their scattered counterparts.

\subsection{Linearity with respect to the radiation field}
\label{sec:linearisation}
In our formulation, the non-linearity of the RT problem in $\bm{I}$ arises from
the frequency-integrated absorption coefficient $k$, 
which enters both the propagation matrix $K$ and the emission coefficient $\pmb{\varepsilon}$.
Indeed, the coefficient $k$ is proportional to the population of the lower level/term, 
which, in turn, depends non-linearly on the incident radiation field through the SE equations.
Notably, the problem becomes linear in $\bm{I}$ if the population of the lower level/term is known a priori 
\citep[see, e.g.][]{janett2021a}.
In this case, the propagation matrix $K$ becomes independent of the radiation field~$\bm{I}$, while $\pmb{\varepsilon}$ depends linearly on $\bm{I}$.
We note that this is formally equivalent to working in optical depth space ($\tau$, see Sect.~\ref{sec:formal_solution}) and incorporating the absorption coefficient, given by the known population of the lower level/term, into the coordinate transformation $\rr\rightarrow\tau$~\citep{janett2021a}. In this case, the RT problem can likewise be formulated as a linear system~\citep[see, e.g.][]{trujillo_bueno1999}.

In the following, we assume that the polarisation of the radiation field has a negligible impact on the population of the lower level/term, which can thus be reliably obtained from independent RT calculations that neglect polarisation and magnetic fields. 
This assumption appears justified, given that we are dealing with resonance lines and that the polarisation of the solar radiation, which rarely exceeds a few percent, should only marginally impact the population of ground (or metastable) levels.
An analogous approach was also used by \citet[][]{faurobert_scholl1992}.
The lower level/term population thus becomes a fixed input parameter of our polarised RT problem, making it linear.
A quantitative verification of the accuracy of this approach is presented in Sect.~\ref{sec:validation_solution}.

We note that already available RT codes, like RH~\citep{uitenbroek2001} and Multi~\citep{Carlsson1986}, allow solving the (unpolarised) NLTE RT problem considering comprehensive multi-level atomic models. 
Such codes can provide very accurate estimates of the lower-level/term population, through which the considered approach can provide more reliable results than those that would be obtained with self-consistent two-level/term atom calculations.

\subsection{Discretisation}
To numerically solve the RT problem introduced in the previous subsections, one has first to discretise the continuum variables of the problem $\rr$, $\vec{\Omega}$, and $\nu$.
The spatial grid points are provided by the considered atmospheric model, which usually discretises the spatial domain through a Cartesian grid with $N_r$ nodes.
The angular discretisation, which samples the unit sphere
with $N_\Omega$ angular nodes,
is usually defined in terms of the quadrature chosen to evaluate the angular integral in Eq.~\eqref{eq:eps_sc_l}, that is, we use the same angular grid for the incident and scattered radiation.
The angular quadrature is thoroughly examined in Sect.~\ref{sec:angular_quadrature}.
Finally, the approximate solution of RT problems
also requires a frequency discretisation suitable to sample
the spectral features present in the Stokes profiles.
Therefore, we discretise the considered finite spectral interval
$[\nu_{\min},\nu_{\max}]\subset \mathbb R_+$ with $N_\nu$ unevenly spaced frequency nodes, which are approximately uniformly spaced in the line core(s)
of the considered spectral line(s), and logarithmically distributed in the wings, where less spectral resolution is needed.

\subsection{Algebraic formulation}
\label{sec:algebraic}

Following a lexicographic criterion ordering \citep[see][]{benedusi2023}, we collect the discretised Stokes parameters in the vector $\bm{I}\in\mathbb R^N$, with $N=4N_rN_\nu N_\Omega$ the total number of degrees of freedom,
which is also the number of unknowns for the considered problem.\footnote{In the following, for the ease of notation, the symbols $\bm{I}$ and $\pmb{\varepsilon}$ are used to represent both continuous and discrete quantities.}
Equation~\eqref{eq:RT} can thus be expressed in the compact matrix form
\begin{equation}
\bm{I}=\Lambda\pmb{\varepsilon}+\bm{t},\label{eq:matrix_form_1}
\end{equation}
where the linear transfer operator $\Lambda\in R^{N\times N}$
encodes the formal solution (see Sect.~\ref{sec:formal_solution})
and the vectors $\pmb{\varepsilon}\in R^N$ and $\bm{t}\in\mathbb R^N$ represent the total emissivity and the radiation transmitted from the boundaries, respectively. 
We note that, using a reverse lexicographic criterion ordering, that is, considering $\Lambda=U^T\tilde\Lambda U$, with $U$ a suitable permutation matrix, one can write the transfer operator as a block diagonal matrix 
$\tilde\Lambda=\mathrm{diag}\left(\tilde\Lambda_{11},\tilde\Lambda_{12},...,\tilde\Lambda_{21},...,\tilde\Lambda_{N_\Omega N_\nu}\right)$,
where the blocks $\tilde\Lambda_{ij}\in\mathbb R^{4N_r\times4N_r}$ (with $i=1,...,N_\Omega$ and $j=1,...,N_\nu$) are lower (for upwards rays) or upper (for downwards rays) triangular because of the structure of the formal solution.
Analogously, we can express the calculation of the emission coefficient as
\begin{equation}
\pmb{\varepsilon}=\Sigma\bm{I}+\pmb{\varepsilon}^{\text{th}},\label{eq:matrix_form_2}
\end{equation}
where the vector $\pmb{\varepsilon}^{\text{th}}\in\mathbb R^N$ represents the thermal contribution to emissivity
and the linear scattering operator
$\Sigma\in R^{N\times N}$ encodes the numerical quadratures necessary to evaluate Eq.~\eqref{eq:eps_sc_l}, respectively.\footnote{We note that the continuum processes contribute to the emissivity with a thermal and a scattering term that are formally analogous to those of line processes (see Appendix~\ref{app:continuum}).}
Here,
$\Sigma=\mathrm{diag}\left(\Sigma_1,...,\Sigma_{N_r}\right)$
is a block diagonal matrix, with the spatially local blocks $\Sigma_i\in\mathbb R^{4N_\nu N_\Omega\times4N_\nu N_\Omega}$ ($i=1,...,N_r$) that are dense because of the structure of $R$ in Eq.~\eqref{eq:eps_sc_l}.
As sparsity patterns suggest, the
application of $\Lambda$ and $\Sigma$
has a very different computational
cost. Indeed, $O(N)$ and $O(NN_\Omega N_\nu)$ floating point operations are required to apply $\Lambda$ and $\Sigma$, respectively.

Equations~\eqref{eq:matrix_form_1} and~\eqref{eq:matrix_form_2}
can be combined into the matrix equation with unknown $\bm{I}$
\begin{equation}\label{eq:linear_system}
(I\hspace{-0.1em}d-\Lambda\Sigma)\bm{I}=\Lambda\pmb{\varepsilon}^{\text{th}}+\bm{t},
\end{equation}
where $I\hspace{-0.1em}d\in\mathbb R^{N\times N}$ is the identity matrix. 
The explicit assembly of the dense coefficient matrix $I\hspace{-0.1em}d - \Lambda\Sigma$ is not suitable for practical applications, because $N$ is typically $\sim\!\!10^7$ for 1D plane-parallel atmospheric models and exceeds $10^{10}$ for full 3D atmospheric models. 
Crucially, we opted for matrix-free applications of the transfer and scattering operators and for an efficient iterative solution to solve the large linear system in Eq.~\eqref{eq:linear_system}.
In particular, an efficient strategy to numerically compute the actions of both $\Lambda$ on $\pmb{\varepsilon}$ and $\Sigma$ on $\bm{I}$ is exposed in Sects.~\ref{sec:formal_solution} and~\ref{sec:calculation_eps}, respectively.
An efficient iterative strategy to solve Eq.~(\ref{eq:linear_system}) for $\bm{I}$, given $\Lambda$, $\Sigma$, $\pmb{\varepsilon}^{\text{th}}$, and $\bm{t}$, is then presented in Sect.~\ref{sec:iterative}.

\section{Formal solution}\label{sec:formal_solution}

The application of $\Lambda$ in Eq.~\eqref{eq:matrix_form_1} commonly requires two steps: 
the conversion from the geometrical scale to the optical depth scale, denoted as $\tau$, and the subsequent numerical integration of Eq.~\eqref{eq:RT} with a suitable formal solver.
The conversion to the optical scale is required both to apply an exponential integrator (i.e. the DELO-methods) and to enforce the numerical stability of the formal solution \citep[e.g.][]{janett2018formal}.
The accuracy of the formal solution is thus impacted by the choice of both the formal solver and the optical depth conversion scheme~\citep{danna2024}.

We note that the RT equation is inherently 1D in space, as it describes how the radiation
varies while propagating along a straight line.
Hereafter, we thus consider the RT problem for the 1D plane-parallel setting, noting that the generalisation to the 3D case is conceptually straightforward.
Formally, in a 1D plane-parallel atmosphere, the coordinate $\rr=(x,y,z)$ is replaced by the vertical coordinate $z \in [z_{\mathrm{min}},z_{\mathrm{max}}]$.
By defining the optical depth scale as the solution of the initial value problem
\begin{equation*}
{\rm d}\tau(z,\vec{\Omega},\nu) = -\frac{\eta_1(z,\vec{\Omega},\nu)}{\cos(\theta)}{{\rm d}z},
\end{equation*}
with the initial condition $\tau(z_{\max},\vec{\Omega},\nu) = 0$ and where $\eta_1\in\mathbb R_+$ is the total absorption coefficient for intensity,
we recast Eq.~\eqref{eq:RT} in the form
\begin{equation}\label{eq:RTE_tau}
\frac{\rm d}{{\rm d} \tau}
 \bm I(\tau,\vec{\Omega},\nu) = 
\frac{K(\tau,\vec{\Omega},\nu)}{\eta_1(\tau,\vec{\Omega},\nu)} 
	\bm I(\tau,\vec{\Omega},\nu) - 
	\frac{\pmb{\varepsilon}
 (\tau,\vec{\Omega},\nu)}{\eta_1(\tau,\vec{\Omega},\nu)},
\end{equation}
where the dependence of $\tau$ on $\vec{\Omega}$ and $\nu$ has not been 
explicitly indicated for notation simplicity.
Here, the term $\pmb{\varepsilon}(\tau,\vec{\Omega},\nu)/\eta_1(\tau,\vec{\Omega},\nu)$ corresponds to the well known source vector.
The increment of optical depth between two generic height nodes 
$z_i$ and $z_{i+1}$ (assuming $z_{i+1}>z_i$, and thus $\theta\in[0,\pi/2)$)
for a given direction $\vec{\Omega}$ is computed by
solving
\begin{equation*}
\Delta \tau(\vec{\Omega},\nu)=
\frac{1}{\cos(\theta)}\int_{z_i}^{z_{i+1}}\!\!\!\eta_1(z',\vec{\Omega},\nu) {\rm d}z'\ge0,
\end{equation*}
which requires a suitable quadrature scheme \citep[e.g. trapezoidal, backward parabolic, spline, cubic Hermite; see][]{danna2024}.

Once the optical depth conversion scheme is set, one can solve Eq.~\eqref{eq:RTE_tau} by means of a suitable integrator, provided that appropriate boundary conditions are given.
Presently, we assume that the radiation entering the atmospheric model from the lower boundary is isotropic, unpolarised, and Planckian in the Wien limit, and that no radiation is entering from the upper boundary.
Finally, given $K$, $\pmb{\varepsilon}$, and $\tau$ at all spatial points, Eq.~\eqref{eq:RTE_tau}
is numerically solved for each individual ray of orientation $\vec{\Omega}$ and frequency $\nu$ by applying a suitable formal solver, such as the DELO-linear, DELOPAR, DELO-parabolic, and BESSER exponential integrators 
\citep[e.g.][]{trujillo_bueno2003,stepan2013,janett2017formal,janett2018formal}.

\section{Calculation of emissivity}\label{sec:calculation_eps}
In this section, we present a strategy to perform an accurate and efficient application of $\Sigma$ in Eq.~\eqref{eq:matrix_form_2}. 
This involves the evaluation of $\pmb{\varepsilon}^{{\ell,\rm sc}}$ through Eq.~\eqref{eq:eps_sc_l}, which requires both an angular quadrature (see Sect.~\ref{sec:angular_quadrature}) and a spectral quadrature (see Sect.~\ref{sec:spectral_quadrature}).
Since the evaluation of $\mathcal R^{\scriptscriptstyle \rm II}$ is generally the computational bottleneck of PRD--AD numerical codes~\citep[e.g.][]{riva2024phd}, our strategy aims at minimising the number of its evaluations, as detailed in the following.

\subsection{Treatment of bulk velocities}
\label{sec:bulkvel}

The numerical solution of RT problems in dynamic atmospheres is more complex than its static counterpart.
Indeed, the presence of a local bulk velocity field $\bm{v}_\mathrm{b}$ introduces frequency (Doppler) shifts in the radiation field, as perceived by the atoms.
These shifts depend on the projections $\bm{v}_\mathrm{b} \cdot \vec{\Omega}'$ (for the absorbed radiation) and $\bm{v}_\mathrm{b} \cdot \vec{\Omega}$ (for the emitted radiation) and add angular and spatial dependencies to the arguments of the frequency-dependent functions appearing in the definitions of both $K$ and~$\pmb{\varepsilon}$.

In order to speed up the computation of $\pmb{\varepsilon}^{{\rm sc}}$, and drawing inspiration from the work by \citet{leenaarts2012}, Eq.~\eqref{eq:eps_sc_l} is evaluated in the comoving reference frame, in which the local bulk velocity is zero.
Namely, we first transform the incident radiation field from the observer's frame to the comoving frame:
\begin{equation*}
    \bm{I}(\rr,\vec{\Omega}',\nu')\;\longrightarrow\;
\bar{\bm{I}}(\rr,\vec{\Omega}',\nu')=
\bm{I}(\rr,\vec{\Omega}',\bar\nu'),
\end{equation*}
with $\bar\nu'=\nu'+\nu_0\bm{v}_\mathrm{b} \cdot\vec{\Omega}'/c$ denoting the incident Doppler shifted frequency
and $\bar{\bm{I}}$ the Stokes vector in the comoving frame.
This is easily performed by means of interpolations.
The scattering integral is then computed in the comoving frame:
\begin{equation}\label{eq:eps_sc_bar}
        \bar{\pmb{\varepsilon}}^{{\rm sc}}(\rr,\vec{\Omega},\nu) \!=\! 
        k(\rr)\!\! \int\!\! {\rm d} \nu'\! 
        \oint\! \frac{{\rm d} \vec{\Omega}'}{4 \pi}
        \bar{R}(\rr,\Theta,\nu',\nu) 
        \bar{\bm{I}}(\rr,\vec{\Omega}',\nu'),
\end{equation}
with $\bar{R}(\rr,\Theta,\nu',\nu)=R(\rr,\vec{\Omega}',
	\vec{\Omega},\bar\nu',\bar\nu)$ the redistribution matrix evaluated in the comoving frame, being $\bar\nu=\nu+\nu_0\bm{v}_\mathrm{b} \cdot\vec{\Omega}/c$.
The advantage of this choice is that $\bar{R}$ does not depend on all the couples of angles~$(\vec{\Omega}',\vec{\Omega})$, but only on the scattering angle $\Theta$~(see Sect.~\ref{sec:redistribution_functions}).
At the same time, the dedicated quadrature grid 
for calculating the scattering integral is independent of the Doppler shifts.
This way, the number of evaluations of the redistribution functions can be drastically reduced (see Sects.~\ref{sec:angular_quadrature} and \ref{sec:spectral_quadrature}). 

Afterwards, the ensuing emission coefficient is transformed from the comoving frame back to the observer's frame, namely
\[\bar{\pmb{\varepsilon}}^{{\rm sc}}(\rr,\vec{\Omega},\nu)\;\longrightarrow\;
\pmb{\varepsilon}^{{\rm sc}}(\rr,\vec{\Omega},\nu)=
\bar{\pmb{\varepsilon}}^{{\rm sc}}(\rr,\vec{\Omega},\nu-\nu_0\bm{v}_\mathrm{b} \cdot\vec{\Omega}/c).\]
Once again, this is performed by means of interpolations,
from the comoving to the observer's frame frequency grid.
We note that interpolations between the $\nu$ and $\bar\nu$ grids allow for the use of different discretisation schemes in the observer's and comoving frames. In this work, we adopt the same grid in both frames for simplicity. Moreover,
the presence of numerical instabilities (e.g. oscillations) in the computed Stokes profiles may indicate that the frequency grid of the problem is not sufficiently dense (or wide) for the considered bulk velocities.
A possible remedy to this issue is the use of
high-order non-oscillatory
interpolations \citep[e.g.][]{janett2019}.

\subsection{Angular quadrature}\label{sec:angular_quadrature}
When selecting the angular quadrature to be applied to Eq.~\eqref{eq:eps_sc_l},
one has to consider some constraints
imposed by the nature of the RT problem \citep[e.g.][]{riva2023}.
First, the number of distinct scattering angle nodes $N_\Theta$ should be kept as small as possible. 
Indeed, when implementing the optimisation strategy
described in Sect.~\ref{sec:bulkvel},
the required number of evaluations of $\mathcal R_Q^{{\scriptscriptstyle \rm II},KK'}$
scales with $N_\Theta$, rather than with $N_\Omega^2$.
Thus, if $N_\Theta\ll N_\Omega^2$, we can achieve 
a significant speed-up of the computation of the emissivity.
Second, considering that the integration of the
scattering emissivity requires a large time-to-solution,
one must guarantee a given accuracy with
the smallest possible number of quadrature nodes.
Third, quadrature nodes that imply a backward scattering geometry (i.e. the scattering angle $\Theta=\pi$) should be avoided
(see Sect.~\ref{sec:spectral_quadrature}).
We finally note that quadrature nodes with an inclination $\theta = \pi/2$ are usually avoided
in practical applications, because horizontal directions never encounter the vertical top and bottom boundaries of the spatial domain.

Many methods for performing the angular
quadrature have been proposed in the literature
\citep[e.g.][]{Carlson1963,koch2004,hesse2015},
also with a focus on the RT modelling of scattering polarisation signals~\citep[e.g.][]{stepan2020,jaumebestard2021quad}.
In particular, \citet{riva2024phd} analysed different angular quadrature rules for the PRD--AD setting, finding that the most suitable ones
are the approaches based on the Cartesian product and Lebedev's rules.
In particular, the Cartesian product rule guarantees a high order of accuracy and easily allows one to avoid both horizontal directions 
(e.g. by using for the inclination a Gauss-Legendre, GL, quadrature with an even number of nodes, or two contiguous GL quadratures)
and the backward scattering geometry 
(e.g. by using for the azimuth a trapezoidal quadrature with an odd number of nodes).
We thus adopted an
angular quadrature given by the tensor product of a trapezoidal quadrature for the azimuthal interval $[0,2\pi)$ for $\chi$
and two GL quadrature nodes, one in each inclination subinterval $[-1,0]$ and $[0,1]$ for the heliocentric angle $\mu=\cos(\theta)$.

\subsection{Spectral quadrature}\label{sec:spectral_quadrature}
Hereafter, it is convenient to work directly with the reduced frequencies $u_n=u(\nu_n)$, where the set $\{\nu_n\}_{n=1}^{N_\nu}$ represents the discrete points of the frequency grid.
The integral over $\nu'$ in Eq.~\eqref{eq:eps_sc_bar} is then numerically evaluated as
\begin{align*}
\int\!\!\bar{R}(\rr,\Theta,\nu'&,\nu_n)
\bar{\bm{I}}(\rr,\vec{\Omega}',\nu'){\rm d}\nu'\approx\\
&\Delta\nu_\mathrm{D}\sum_{n'=1}^{N_{\nu'}}w_{n'}\bar{R}(\rr,\Theta,u_{n'},u_n) 
\bm{I}\left(\rr,\vec{\Omega}',{u}_{n'}-u_\mathrm{b}(\vec{\Omega}')\right),
\end{align*}
where $\{u_{n'}\}_{n'=1}^{N_{\nu'}}$ and $\{w_{n'}\}_{n'=1}^{N_{\nu'}}$ are the chosen quadrature nodes and weights, respectively.
Here, we used 
$\bm{I}(\rr,\vec{\Omega},\nu)\mathrm{d}\nu=\Delta\nu_\mathrm{D}\bm{I}(\rr,\vec{\Omega},u)\mathrm{d}u$, 
noting that, in this work, we always define
the Stokes parameters $\bm{I}$, as well as the redistribution matrices $R$, per unit of frequency.
We also note that, to apply the chosen quadrature rule, the radiation field $\bm{I}$ has to be interpolated from the original grid $u_n$ to the quadrature grid in the comoving frame ${u}_{n'}-u_\mathrm{b}(\vec{\Omega}')$.
Moreover, since the redistribution functions $\mathcal{\bar R}^{\scriptscriptstyle \rm II}$ and $\mathcal{\bar R}^{\scriptscriptstyle \rm III}$ (the overbars indicating that they are expressed in the comoving frame, as discussed in Sect.~\ref{sec:bulkvel}) display very different computational complexities,
distinct quadrature grids 
are used to evaluate the integral over $\nu'$ for the two different cases.
These are detailed below.

Given the high computational cost of evaluating $\bar{R}^{\scriptscriptstyle \rm II}$,
it is crucial to minimise $N_{\nu'}$.
By considering that the transformation of $G$ from the observer's to the comoving frame is
\begin{align*}
G(\vec{\Omega}',\vec{\Omega},\nu_{n'},\nu_n,k_u,k_\ell,k_\ell')\;\longrightarrow\;&
\bar{G}(\Theta,\nu_{n'},\nu_n,k_u,k_\ell,k_\ell')\\
=&F\left(\Theta,u_{n'}+u_{k_u,k_\ell},u_n+u_{k_u,k_\ell'}\right),
\end{align*}
our selection of $\{u_{n'}\}_{n'=1}^{N_{\nu'}}$ and $\{w_{n'}\}_{n'=1}^{N_{\nu'}}$ is based on the following properties:
(a)~the behaviour in frequency of the integrands in Eq.~\eqref{eq:eps_sc_bar} is mostly driven by the dependence of $\bar{R}$ on $\Theta$, $\nu'$, and $\nu$, and only
to a lesser extent by the incident radiation field;
(b)~for typical solar chromospheric conditions, $a\ll1$;
(c)~the function $F$ quickly drops to~0 as $|u_{n'}-u_n-u_{k_\ell,k_\ell'}|$ increases, 
with $u_{k_\ell,k_\ell'}=\nu_{k_\ell,k_\ell'}/\Delta\nu_\mathrm{D}$ the reduced frequency splitting between states pertaining to the same level or term;
(d)~if $\Theta$ is small or $|u_n+u_{k_u,k_\ell'}|$ is large, then $F$ is close to a Gaussian function centred on $u_{n'}=u_n+u_{k_\ell,k_\ell'}$;
(e)~if $|u_n+u_{k_u,k_\ell'}|\lesssim2$, then $\mathrm{Re}(F)$ has one local maximum and $\mathrm{Im}(F)$ a sharp sign reversal close to $u_{n'}=(u_n+u_{k_u,k_\ell'})\cos(\Theta)-u_{k_u,k_\ell}$, 
with $\mathrm{Re}$ and $\mathrm{Im}$ denoting the real and the imaginary part, respectively;
(f)~if $2\lesssim|u_n+u_{k_u,k_\ell'}|\lesssim 6$, then $\mathrm{Re}(F)$ displays two local maxima and $\mathrm{Im}(F)$ one local extremum and a sharp sign reversal, respectively;
(g)~the case $\Theta=\pi$ introduces extremely sharp peaks that are difficult to integrate numerically.
The regimes~(e) and~(f) are generally referred to as line core and near wings, respectively.
\begin{figure*}[ht!]
    \centering
    \includegraphics[width=0.95\textwidth]{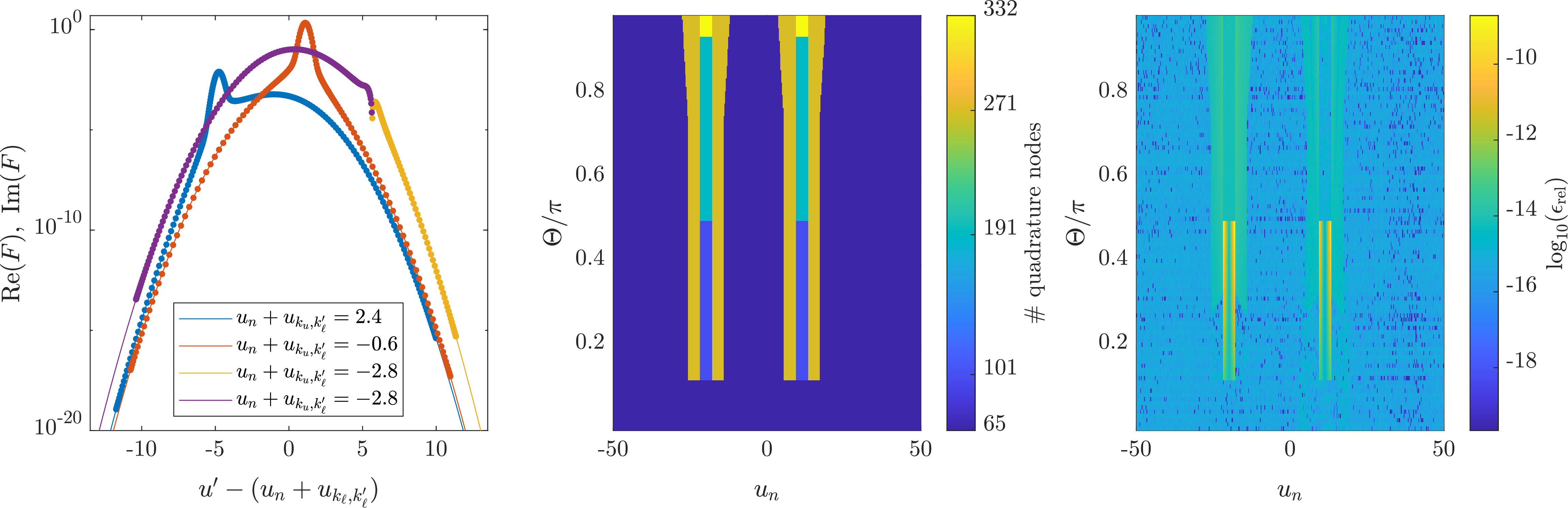}
    \caption{Illustrative examples of the quadrature grids. Left: real part (blue and red lines) and positive and negative values of the imaginary part~(yellow and purple lines, respectively) of $F$ as a function of 
    $u'-(u_n+u_{k_\ell,k_\ell'})$, with $u'=u(\nu')$, $u_n=u(\nu_n)$, and $u_{k_\ell,k_\ell'}=u_{k_u,k_\ell'}-u_{k_u,k_\ell}$, for different values of $u+u_{k_u,k_\ell'}$. Here, $a=0.01$, $\Theta=0.9\pi$, $u_{k_u,k_\ell}=-11.1$, and $u_{k_u,k_\ell'}=-11.2$. The dots denote the quadrature nodes $u_{n'}$ generated with $u^*=-11.15$. 
    Centre: number of quadrature nodes as function of $u_n$ and $\Theta$, with $a=0.01$, $u_1^*=-11.15$, and $u_2^*=20$. 
    Right: logarithm of the relative error of evaluating Eq.~\eqref{eq:int_Ftilde} with the quadrature nodes of the central panel with respect to the reference values obtained with the Gauss–Kronrod adaptive method, with $\bm{I}=(1,0,0,0)$, $a=0.01$, and $u_{k_u,k_\ell}=u_{k_u,k_\ell'}=-11.1$.}
    \label{fig:nu_quadrature_nodes}
\end{figure*}
For instance, the left panel of Fig.~\ref{fig:nu_quadrature_nodes} displays $\mathrm{Re}(F)$ in the near-wing (blue line) and line-core (red line) regimes, as well as the positive (yellow line) and negative (purple line) values of $\mathrm{Im}(F)$ in the near-wing regime. 
We refer to Chapter 2 of~\citet{riva2024phd} for a more comprehensive analysis of the behaviour of the redistribution function $\mathcal R^{\scriptscriptstyle \rm II}$.
We finally note that, under typical solar chromospheric conditions, the shifts $u_{k_u,k_\ell}$ cluster around either one single value (for either a two-level atom or a two-term atom with completely blended fine-structure components, such as H~{\sc i} Ly-$\alpha$) or around a few values (e.g. two for the Mg~{\sc ii} h\&k doublet), with the distance of $u_{k_u,k_\ell}$ from the centre of the corresponding cluster being much smaller than one. 
We denote the centre of each cluster as $u_i^*$, with $i=1,\ldots,\#$ of clusters.\footnote{If the distance between the centres of two clusters is smaller than 0.25, we consider the two clusters as a single one.}

In practice, we do not need to generate a quadrature grid for each possible combination $(k_u,k_\ell,k_\ell')$. Instead, it is enough to consider a set of quadrature nodes that well approximates the integral
$\int\!F(\Theta,u'+u_i^*,u_n+u_j^*)\mathrm{d}u'$ for all couples $(i,j)$. This means that the selection of $\{u_{n'}\}_{n'=1}^{N_{\nu'}}$ only depends on $\rr$, $\Theta$, $u_n$, and $u_i^*$.\footnote{We recall that $F$ depends on $a$, which in turn is a function of $\rr$, and that also $u_n$ depends on $\rr$ through the $\Delta\nu_\mathrm{D}$ term.}
This property of the spectral quadrature allows us to reduce the number of interpolations of $\bm{I}$ required to evaluate the emissivity.

The strategy adopted in this work, which is detailed in Appendix~\ref{app:spectral_quadratrue}, can be summarised as follows. 
If $\Theta=0$, then $F(0,u,u')=\sqrt{\pi}\,\delta(u-u')W(a,u')$ and the integral 
\begin{equation}\label{eq:int_Ftilde}
\int\!F(\Theta,u'+u_{k_u,k_\ell},u_n+u_{k_u,k_\ell'})\bm{I}\left(\rr,\vec{\Omega}',u'-u_\mathrm{b}(\vec{\Omega}')\right)\mathrm{d}u'    
\end{equation}
is carried out analytically. Otherwise, if $\Theta$ is small or $-u_n$ is far from all $u_i^*$, we use a Gauss-Hermite quadrature with 65 nodes, thus exploiting property (d). In all other cases, we are either in the line-core~(e) or in the near-wing~(f) regimes, and we adopt a GL quadrature, with a number of nodes that depends on $\Theta$ and $|u_n+u_i^*|$.
An additional strategy adopted in this work to further diminish the number of evaluations of the Faddeeva function is presented in Appendix~\ref{app:strategy}.

As illustrative examples, the left panel of Fig.~\ref{fig:nu_quadrature_nodes} shows the frequency quadrature nodes as dots for the case of a single cluster with $u^\ast=-11.15$. The central panel displays the number of frequency nodes $N_{\nu'}$ generated with $u_1^*=-11.15$ and $u_2^*=20$ for different $u_n$ and $\Theta$. Finally, the right panel illustrates the relative error of evaluating Eq.~\eqref{eq:int_Ftilde} with the strategy outlined above, with $\bm{I}=(1,0,0,0)$, $a=0.01$, $u_{k_u,k_\ell}=u_{k_u,k_\ell'}=-11.1$, $u_1^*=-11.15$, and $u_2^*=20$.\footnote{These numerical values are just to exemplify the grids used in the present work.} Notably, with about $100-300$ spectral quadrature nodes, we obtain a relative error that is always smaller than $10^{-8}$.

Concerning the $\mathcal{\bar R}^{{\scriptscriptstyle \rm III}}$ redistribution function, the situation is much simpler than for its $\mathcal{\bar R}^{{\scriptscriptstyle \rm II}}$ counterpart, as the incident and emergent frequencies are totally uncorrelated~(see Sect.~\ref{sec:redistribution_functions}).
Therefore, for the numerical integration of $\mathcal{\bar R}^{\scriptscriptstyle \rm III}\cdot I$ over frequency, it is possible to choose a quadrature grid independent of $\rr$, $\Theta$, and $u_n$. In the following, we consider a single GL rule with 2001 nodes in the interval $[-200,200]$.

We note that, when evaluated numerically, Eq.~\eqref{eq:eps_sc_l}
should be suitably normalised to avoid
spurious sources or sinks of photons~\citep[e.g.][]{adams1971}.
In Appendix~\ref{app:normalisation}, we show how the generalised Kirchhoff's law can be used
to perform this task.
This method is presently used to normalise the intensity component of the emission vector when computing the results presented in Sect.~\ref{sec:results}.
By applying this procedure, we improved both the accuracy of the results and the robustness of the iterative method.

\section{Iterative solution strategy}\label{sec:iterative}
The considered RT problem has been recast
in the linear system given by Eq.~\eqref{eq:linear_system},
for which we have to find a solution.
Given the size of the system ($N \sim 10^7$ for small-scale 1D applications, see Sect.~\ref{sec:num_par}) and that the matrix $I\hspace{-0.1em}d-\Lambda\Sigma$ in Eq.~\eqref{eq:linear_system} is dense, the use of direct matrix inversion methods
is unfeasible. 
Thus, the application of an iterative solution strategy is necessary.

\subsection{Matrix-free iterative methods}
Krylov methods are well suited for matrix-free approaches and proved to be highly effective solution
strategies in the RT context for large linear systems \citep[e.g.][]{jolivet2021,benedusi2021,benedusi2023}.
In the present work, we thus consider the generalised minimal residual (GMRES) iterative method in its matrix-free version.
For increased flexibility, we also explore other iterative solution strategies, such as the Richardson and the alternating Anderson-Richardson (AAR) methods~\citep{hackbusch1994iterative,AAR_2019}.
We monitored the convergence of the iterative methods by evaluating the relative residual
\begin{equation*}
res = \frac{\|
\Lambda\pmb{\varepsilon}^{\text{th}}+\bm{t}-
(Id-\Lambda\Sigma)\bm{I}\|_2}{\|\Lambda\pmb{\varepsilon}^{\text{th}}+\bm{t}\|_2},
\end{equation*}
taking as a stopping criterion $res<tol$,
for a given tolerance $tol$.

\subsection{Preconditioning}\label{sec:preconditioning}
Preconditioning of linear systems
can significantly increase robustness and convergence speed of
iterative techniques.
Being $P\in\mathbb R^{N\times N}$ a non-singular matrix,
left preconditioning\footnote{We consider only left preconditioning
because no extra step is required to compute the final solution.} of Eq.~\eqref{eq:linear_system}
is obtained by considering the equivalent system
\begin{equation*}
P^{-1}(I\hspace{-0.1em}d-\Lambda\Sigma)\bm{I}
=P^{-1}(\Lambda\pmb{\varepsilon}^{\text{th}}+\bm{t}).
\end{equation*}
We remind that, to be effective, $P$
should be a good and computationally cheap approximation of the operator
$I\hspace{-0.1em}d - \Lambda\Sigma$.

\citet{janett2024} showed that, when dealing with 
RT problems for polarised radiation including AD PRD effects, common algebraic preconditioners, such as Jacobi and SOR,
are outperformed by tailored physics-based preconditioners, that is, operators corresponding to a simplified physical setting of the considered problem.
Crucially, physics-based preconditioners do not require the explicit algebraic description of the 
operators of interest and are very intuitive to implement.
In our RT problem, we can write physics-based preconditioners as
\begin{equation*}
P = I\hspace{-0.1em}d - \Lambda^{\ast} \Sigma^{\ast},
\end{equation*}
where $\Lambda^{\ast}$ and $\Sigma^\ast$
encode the approximations of
$\Lambda$ and $\Sigma$.

A variety of simplifications can be adopted to design $\Sigma^{\ast}$. 
For instance, a common approach to lower the computational complexity of $\Sigma$ is to consider the so-called AA approximation~\citep[e.g.][]{mihalas1978,rees1982,bommier1997b}. 
Within this approximation, the redistribution function in the comoving frame describing scattering processes that are coherent in frequency in the atomic rest frame can be written as
\begin{equation}\label{eq:R_AA}
    \mathcal{\bar R}_Q^{{\scriptscriptstyle \rm II-AA},KK'}(\nu',\nu)=
    \frac{1}{2}\int_0^\pi\sin(\Theta)
    \mathcal{\bar R}_Q^{{\scriptscriptstyle \rm II},KK'}(\Theta,\nu',\nu){\rm d} {\Theta}.
\end{equation}
Remarkably, this redistribution function is independent of propagation directions and only couples the frequencies of the incident and scattered radiation.
We note that the AA approximation is not only useful in the context of physics-based preconditioners, but it also yields reliable results when applied to the modelling of the intensity spectra of strong resonance lines \citep[see][]{leenaarts2012,sukhorukov2017}.
As far as scattering polarisation is concerned, the AA approximation can either provide satisfactory results~\citep[e.g.][]{riva24,delpinoaleman2024}, or 
introduce large and hardly-predictable inaccuracies~\citep[e.g.][]{sampoorna2017,janett2021b,belluzzi2024}.
For this reason, it is always important to carefully assess its suitability under different conditions, as discussed in Sect.~\ref{sec:AA_AD_HI} for the H~{\sc i} Ly-$\alpha$ line.

To further simplify $\Lambda^{\ast}$ and $\Sigma^\ast$, it is also possible to
neglect polarisation and magnetic fields, or assume an isotropic emission vector in the comoving frame~\citep{benedusi2022,janett2024}.
We finally note that the application of $P^{-1}$ requires itself solving a linear system of the form $P\bm{x}=\bm{y}$.
This is presently achieved iteratively with a matrix-free iterative method,
meaning that the whole linear problem is solved by means of two nested iterative solvers. To further enhance the performance of our solution strategy, we also implemented the possibility of preconditioning the linear system $P\bm{x}=\bm{y}$, as detailed in Appendix~\ref{app:Block_jacobi}.

\section{Numerical applications}\label{sec:results}
The solution strategy outlined in Sects.~\ref{sec:problem}-\ref{sec:iterative} was implemented in TRAP$^4$ as a MATLAB~\citep{MATLAB:2023} routine, considering plane-parallel 1D models, and parallelised by spatially decomposing the calculation of $\pmb{\varepsilon}^{{\rm sc}}$ across threads.
Hereafter, we discuss concrete applications of TRAP$^4$.
Specifically, after presenting the atmospheric and atomic models and the numerical setting, 
we investigate the numerical error affecting our results, 
we perform a benchmark with the HanleRT-TIC\footnote{The code is publicly available at
\url{https://gitlab.com/TdPA/hanlert-tic}.} code \citep{delpinoaleman2016,delpinoaleman2020},
and we discuss the suitability of using the PRD--AA approximation with respect to full AD calculations. 
For our applications, we considered the following chromospheric resonance lines
whose modelling requires the inclusion of both PRD effects and $J$-state interference:
the Mg~{\sc ii} h\&k doublet at 2800\,{\AA} and the H~{\sc i} Ly-$\alpha$ line at 1216~\AA.
For sake of conciseness, in the following we do not discuss Stokes $V$,
although it is always included in our calculations.

\subsection{Atmospheric and atomic models}
\label{sec:input_parameters}
In this work, calculations were carried out in
a 1D plane-parallel domain extracted from the
semi-empirical atmospheric model C (hereafter FAL-C) of \citet{fontenla1993},
which fully contains the formation region of the considered spectral lines.
Even if 1D, this model is sophisticated enough to test our strategy and assess the suitability of 
using the AA approximation to model wavelength-integrated scattering polarisation signals of the H~{\sc i} Ly-$\alpha$ line.

In the following, we consider the contribution to the Doppler width due to non-thermal microturbulent velocity
\begin{equation*}
\Delta\nu_\mathrm{D}(z)=\frac{\nu_0}{c}\sqrt{\frac{2k_\mathrm{B}T(z)}{M}+\xi(z)^2},
\end{equation*}
where $T$ is the temperature, $M$ the mass of the atom, and $\xi$ the microturbulent velocity determined according to \citet{fontenla1991}.
Moreover, for the sake of generality, some calculations were carried out in the presence of
magnetic and bulk velocity fields, which were added ad-hoc to the atmospheric model, as detailed in the following subsections.

The Mg~{\sc ii} h\&k lines result from two distinct fine-structure transitions between the ground level of ionised magnesium $3s \, ^2\mathrm{S}_{1/2}$
and the first excited term composed by the levels
$3p \, ^2\mathrm{P}_{1/2}^\mathrm{\, o}$ and 
$3p \, ^2\mathrm{P}_{3/2}^\mathrm{\, o}$.
\begin{table*}
\caption{Atomic parameters for the Mg~{\sc ii} h\&k doublet and the H~{\sc i} Ly-$\alpha$ line.}
\label{table:atomic_models}
\centering
\begin{tabular}{c c c c c c c c}
\hline\hline
Ion & $M$ (AU) & Ground state & Line & Upper state & $E_u\ (\mathrm{cm}^{-1})$ & $\lambda\ (\AA)$ & $A_{u\ell}\ (\mathrm{s}^{-1})$\\
\hline
    \multirow{2}{*}{Mg~{\sc ii}} & \multirow{2}{*}{24.305} & \multirow{2}{*}{$3s \, ^2\mathrm{S}_{1/2}$} & 
    h & $3p \, ^2\mathrm{P}_{1/2}^\mathrm{\, o}$ & 35669.31 & 2803.531 & $2.57\times10^8$\\
    \cline{4-8} & & &
    k & $3p \, ^2\mathrm{P}_{3/2}^\mathrm{\, o}$ & 35760.88 & 2796.352 & $2.60\times10^8$\\
    \hline
    \multirow{2}{*}{H~{\sc i}} & \multirow{2}{*}{1.008} & \multirow{2}{*}{$1s \, ^2\mathrm{S}_{1/2}$} &
    \multirow{2}{*}{Ly-$\alpha$} & 
    $2p \, ^2\mathrm{P}_{1/2}^\mathrm{\, o}$ & 
    82258.92 & 1215.674 & $6.265\times10^8$\\
    \cline{5-8} & & & &
    $2p \, ^2\mathrm{P}_{3/2}^\mathrm{\, o}$ &
    82259.29 & 1215.668 & $6.265\times10^8$\\
\hline
\end{tabular}
\tablefoot{Data from~\citet{NIST_ASD}. The quantities $E_u$ and $\lambda$ denote the energy of the upper state and the line-centre wavelength in vacuum, respectively. For both ions, the energy of the ground state is $E_\ell=0$.
The Einstein coefficients for spontaneous emission from the upper to the lower term of the considered atomic models are obtained as the average of the experimental Einstein coefficients for the fine-structure components of each multiplet, weighted by the statistical weights of the corresponding upper $J$-levels~\citep[see Eq.~(1) in ][]{delpinoaleman2020}. For the Mg~{\sc ii} h\&k doublet, we thus used $A_{u\ell}=2.59\times10^8\ \mathrm{s}^{-1}$, whereas for the H~{\sc i} Ly-$\alpha$ line we used $A_{u\ell}=6.265\times10^8\ \mathrm{s}^{-1}$.}
\end{table*}
These two fine-structure components are separated by 7\,{\AA}
and give thus rise to two well distinct lines.
The H~{\sc i} Ly-$\alpha$ line also results from two distinct fine-structure transitions, between the ground level of hydrogen $1s \, ^2\mathrm{S}_{1/2}$
and the excited term composed by the levels
$2p \, ^2\mathrm{P}_{1/2}^\mathrm{\, o}$ and 
$2p \, ^2\mathrm{P}_{3/2}^\mathrm{\, o}$.\footnote{We neglect the contribution from the forbidden transition between the ground level and the excited level 
$2s \, ^2\mathrm{S}_{1/2}$.}
Contrary to the Mg~{\sc ii} h\&k lines, these two fine-structure components are separated by 6\,m{\AA} only,
and are thus completely blended.
The relevant atomic parameters for
the Mg~{\sc ii} h\&k doublet and the
H~{\sc i} Ly-$\alpha$ line are provided in Table~\ref{table:atomic_models}. 

In addition to the atmospheric and atomic parameters, TRAP$^4$ also requires as input quantities the 
broadening constants $\Gamma_{\rm R}$, $\Gamma_{\rm I}$, $\Gamma_{\rm E}$, and the lower-level/term population of the considered transition.
For all our calculations, we used $\Gamma_{\rm R}=A_{u\ell}$, with $A_{u\ell}$ the Einstein coefficient for spontaneous emission from the upper to the lower term (see Table~\ref{table:atomic_models}).
For Mg~{\sc ii}, we calculated the ground-level population using the HanleRT-TIC code, neglecting polarisation and considering a static unmagnetised atmosphere and a two-term atom with AD PRD effects.
The same code was also used to calculate the $\Gamma_{\rm E}$ and $\Gamma_{\rm I}$ broadenings, as well as the continuum coefficients $\kappa^c$, $\sigma^c$, and $\pmb{\varepsilon}^{c,\text{th}}$ (see Appendix~\ref{app:continuum}).
In particular, $\Gamma_{\rm E}$ accounts for elastic collisions with both neutral hydrogen and helium perturbers via Van der Waals interactions and the quadratic Stark effect, while $\Gamma_{\rm I}$ is calculated treating inelastic collisions in a two-term atom following \citet{belluzzi13}.
For H~{\sc i} Ly-$\alpha$, we used the hydrogen ground-level population tabulated in the FAL-C model.
The broadening due to inelastic de-exciting collisions was evaluated as $\Gamma_{\rm I}=N_eQ_{ul}$, with $N_e$ the electron density from the FAL-C model and $$Q_{ul}(z)=8.63\cdot10^{-6}\frac{\upsilon(z)}{g_u\sqrt{T(z)}},$$ where $g_u=8$ is the degeneracy of the upper level with principal quantum number $n=2$ and the coefficient $\upsilon$ (which depends on $T$) is taken according to~\citet{Przybilla2004}.
Furthermore, $\Gamma_{\rm E}$ was computed with the RH code~\citep{uitenbroek2001}, fixing the hydrogen ground-level population to the value provided by the FAL-C model and considering collisions with neutral hydrogen and helium perturbers via both linear and quadratic Stark effects.
The values of $\kappa^c$, $\sigma^c$, and $\pmb{\varepsilon}^{c,\text{th}}$ were also computed with RH.
A similar procedure was used to initialise the H~{\sc i} Ly-$\alpha$ calculations in~\citet[][]{alsinaballester2022}.

\subsection{Physical and numerical parameters}
\label{sec:num_par}
Concerning the spectral grid, we used 210 wavelength nodes in the interval $[2705.5\ \AA,2901.2\ \AA]$ for the Mg~{\sc ii} h\&k doublet and 191 wavelength nodes in the interval $[1184.5\ \AA,1248.6\ \AA]$ for the H~{\sc i} Ly-$\alpha$ line.
For the angular discretisation, we used nine nodes uniformly distributed in $\chi$, 
whereas we used two sets (one for each hemisphere) of six GL nodes for the heliocentric angles $\mu=\cos(\theta)$, for a total $N_\Omega=108$.
Given that $N_r=70$ for the FAL-C atmospheric model, the RT problem in Eq.~\eqref{eq:linear_system} thus had $N=4N_rN_\nu N_\Omega\!\sim6\cdot10^6$ degrees of freedom for both atomic models.
We note that the chosen grids correspond to a number of scattering angles $N_\Theta=205\ll N_\Omega^2=11664$ and an averaged number of spectral quadrature nodes $N_{\nu'}\simeq133$ and $N_{\nu'}\simeq127$ for Mg~{\sc ii} and H~{\sc i}, respectively.
Presently, all interpolation between the observer's and the comoving frame (and vice versa) are performed in terms of cubic splines.

Concerning the formal solver, we used the BESSER exponential integrator for the Mg~{\sc ii}, consistently with the recent work
by~\citet{delpinoaleman2024}. On the other hand, we used the DELOPAR exponential integrator for the H~{\sc i} Ly-$\alpha$ line, to allow direct comparisons with previous works in the literature~\citep[e.g.][]{alsinaballester2022}.
As the BESSER integrator introduces some small non-linearity into the problem due to the source vector interpolation used to avoid overshooting~\citep[see the definition of the control points in][]{stepan2013}, we opted for the AAR iterative method with $tol=10^{-9}$ for the Mg~{\sc ii} h\&k doublet, whereas we used GMRES with $tol=10^{-10}$ for the H~{\sc i} Ly-$\alpha$ line. 
After reaching the desired tolerance, we performed an additional formal solution to compute the final results at specific directions that are not part of the angular grid.

With the above numerical parameters, the code employed approximately 1700\,s for the Mg~{\sc ii} h\&k doublet and 1450\,s for the H~{\sc i} Ly-$\alpha$ line to perform a single preconditioned iteration on an Intel Xeon CPU E5-1620. 
Overall, for both atomic models, the largest fraction of the computational time was spent for evaluating $\pmb{\varepsilon}^{\ell,\text{sc}}$, which took about 1300\,s per iteration. 
Thanks to our solution strategy, only $25\%$ of this time was spent in evaluating the Faddeeva function, whereas approximately $55-60\%$ of it was employed to interpolate $\bm{I}$ to the spectral quadrature grid and assemble $\vec{\mathcal W}_{k_u,k_\ell}$ in Eq.~\eqref{eq:Wkukl}. 
Moreover, by employing the block-diagonal preconditioner introduced in Appendix~\ref{app:Block_jacobi}, we ensured that less than $20\%$ of the full solution time was spent in preconditioning the system.
Finally, given the high performances of physics-based preconditioned methods, the target tolerance was reached with only 10-12 iterations when starting from a zero initial guess.

\subsection{Solution verification}
Before using any numerical results to perform a physical investigation, it is essential to assess if they are accurate enough for their intended use.
This task is known as solution verification.
In this respect, we note that the tolerances specified in Sect.~\ref{sec:num_par} are very conservative, thus ensuring unnoticeable differences in the emerging profiles when further reducing $tol$ by an order of magnitude.
\begin{figure*}[ht!]
    \centering
    \includegraphics[width=0.95\textwidth]{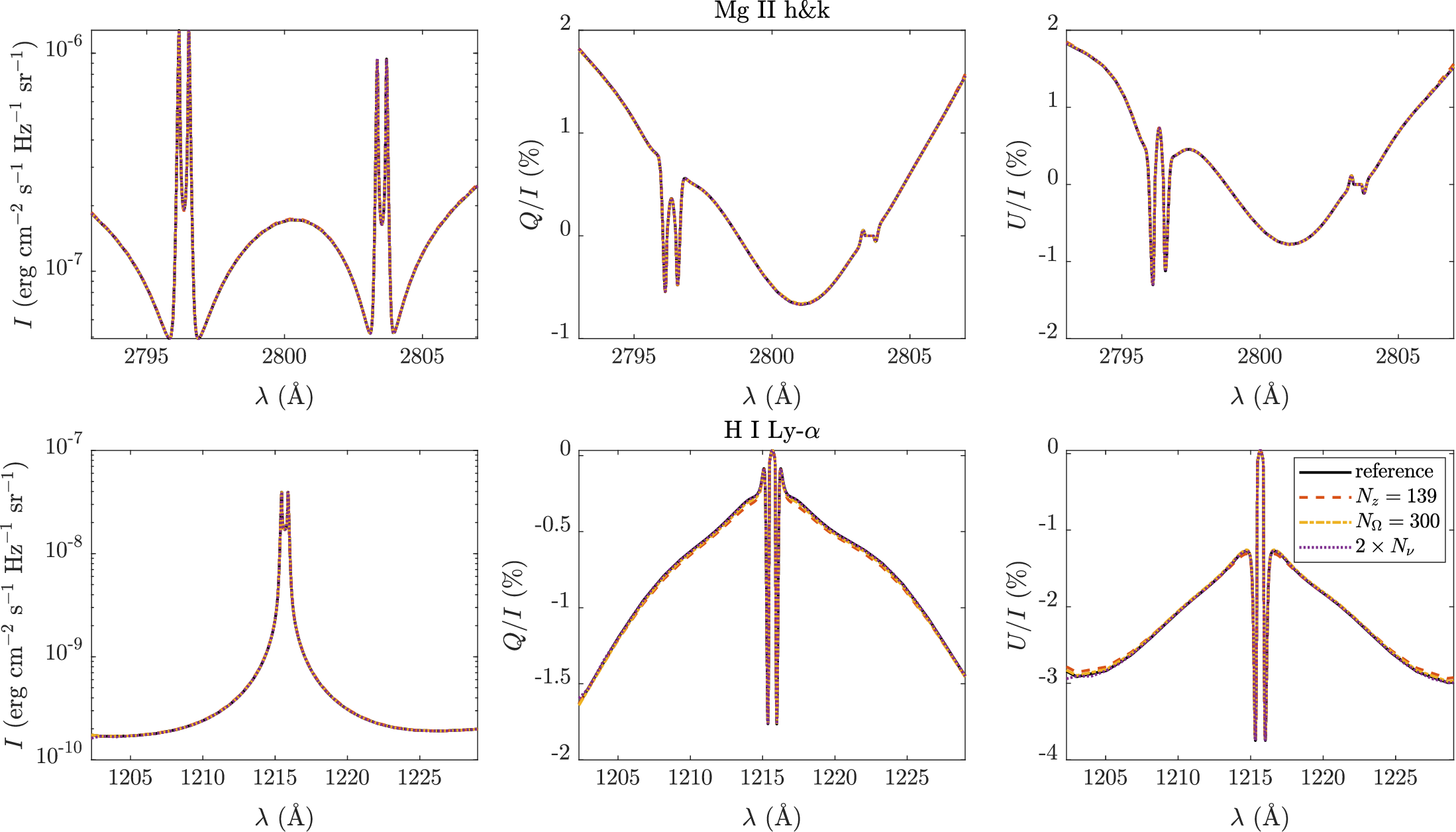}
    \caption{Intensity (left column), $Q/I$ (middle column), and $U/I$ (right column) as a function of vacuum wavelength for the Mg~{\sc ii} h\&k doublet (first row) and the H~{\sc i} Ly-$\alpha$ line (second row), obtained with PRD--AD calculations in a static magnetic FAL-C atmosphere (see text) for a LOS with $\mu=0.1$. The solid black curves correspond to reference calculations, whereas the red (dashed), yellow (dot-dashed), and purple (dotted) lines are the results of calculations with increased spatial, angular, and frequency resolution, respectively. The reference direction for positive Stokes $Q$ is the parallel to the limb.}
    \label{fig:sol_ver}
\end{figure*}
Moreover, besides carrying out the reference calculations with numerical parameters as given in Sect.~\ref{sec:num_par}, we carried out three additional calculations for both the Mg~{\sc ii} h\&k doublet and the H~{\sc i} Ly-$\alpha$ line: 
one with doubled spatial resolution, where we introduced a new set of grid points equidistant to each pair of adjacent original points and interpolated the input variables to the new grid; 
one with increased angular resolution, with fifteen azimuths and two sets of ten GL inclinations for a total $N_\Omega=300$; 
and one with doubled $N_\nu$.
For all these calculations, we considered a static atmosphere with a height-independent horizontal~($\chi_B=0$ and $\theta_B=\pi/2$) magnetic field of $20\ \mathrm{G}$ and $50\ \mathrm{G}$ for the Mg~{\sc ii} h\&k doublet and the H~{\sc i} Ly-$\alpha$ line, respectively.

The resulting profiles for a LOS $\mu=0.1$ are shown in Fig.~\ref{fig:sol_ver}.
In general, we see an excellent agreement between the black solid (reference calculations) and the colour lines.\footnote{Excellent agreement is found also for other LOSs.}
Minor discrepancies are observed away from the line cores, in particular in the fractional linear polarisation signals of the H~{\sc i} Ly-$\alpha$ line.
These extremely small discrepancies suggest that the reference calculations are at convergence in resolution.
Moreover, the profiles look very similar to other computations in the literature~\citep[compare to Figs. 3 and B.3 of][for the Mg~{\sc ii} h\&k doublet and the H~{\sc i} Ly-$\alpha$ line, respectively]{alsinaballester2022}, thus increasing the reliability of our results.
We also carried out an additional set of computations (not shown in Fig.~\ref{fig:sol_ver} for clarity),
where we used the highly accurate (though more time-consuming) adaptive Gauss–Kronrod method for integrating Eq.~\eqref{eq:Wkukl} over frequency.
Negligible differences with respect to the reference emergent profiles were found. 

\subsection{Benchmark with the HanleRT-TIC code}
\label{sec:validation_HanleRT}
To verify the TRAP$^4$ code,
we performed a benchmark study for the Mg~{\sc ii} h\&k doublet with the HanleRT-TIC code.
We recall that the two codes are based on different theoretical frameworks, so that the form of the redistribution matrices $R^{\scriptscriptstyle \rm II}$ and $R^{\scriptscriptstyle \rm III}$ is not exactly the same.
Moreover, they use different frequency grids, quadrature grids, and iterative solution methods.
Also the way of normalising the emissivity is different.

For this test, we considered a height-independent horizontal~($\chi_B=0$ and $\theta_B=\pi/2$) magnetic field of $20\ \mathrm{G}$.
Moreover, to increase the complexity of the problem, we added a bulk velocity to the atmospheric model, with the vertical component mimicking Fig.~1 of~\citet{delpinoaleman2024} and defined as
\begin{equation*}
v_z=-10\cdot e^{-\left[(z-2100)/500\right]^2}\cdot\left[1+\mathrm{erf}\left(\frac{2100-z}{150}\right)\right],
\end{equation*}
the horizontal components given by
\begin{equation*}
v_x = 0.7\,|v_z|\cos(\chi_v)\qquad \mathrm{and} \qquad v_y = 0.7\,|v_z|\sin(\chi_v),
\end{equation*}
and 
\begin{equation*}
\chi_v=-2\pi\frac{z-z_\mathrm{min}}{z_\mathrm{max}-z_\mathrm{min}}+\pi,
\end{equation*}
being $z_\mathrm{min}=-100\ \rm{km}$ and $z_\mathrm{max}=2219.43\ \rm{km}$ and $z$ given in~km.
We then solved the RT problem with the TRAP$^4$ and HanleRT-TIC codes,
for both AA and AD cases.
\begin{figure*}[ht!]
    \centering
    \includegraphics[width=0.95\textwidth]{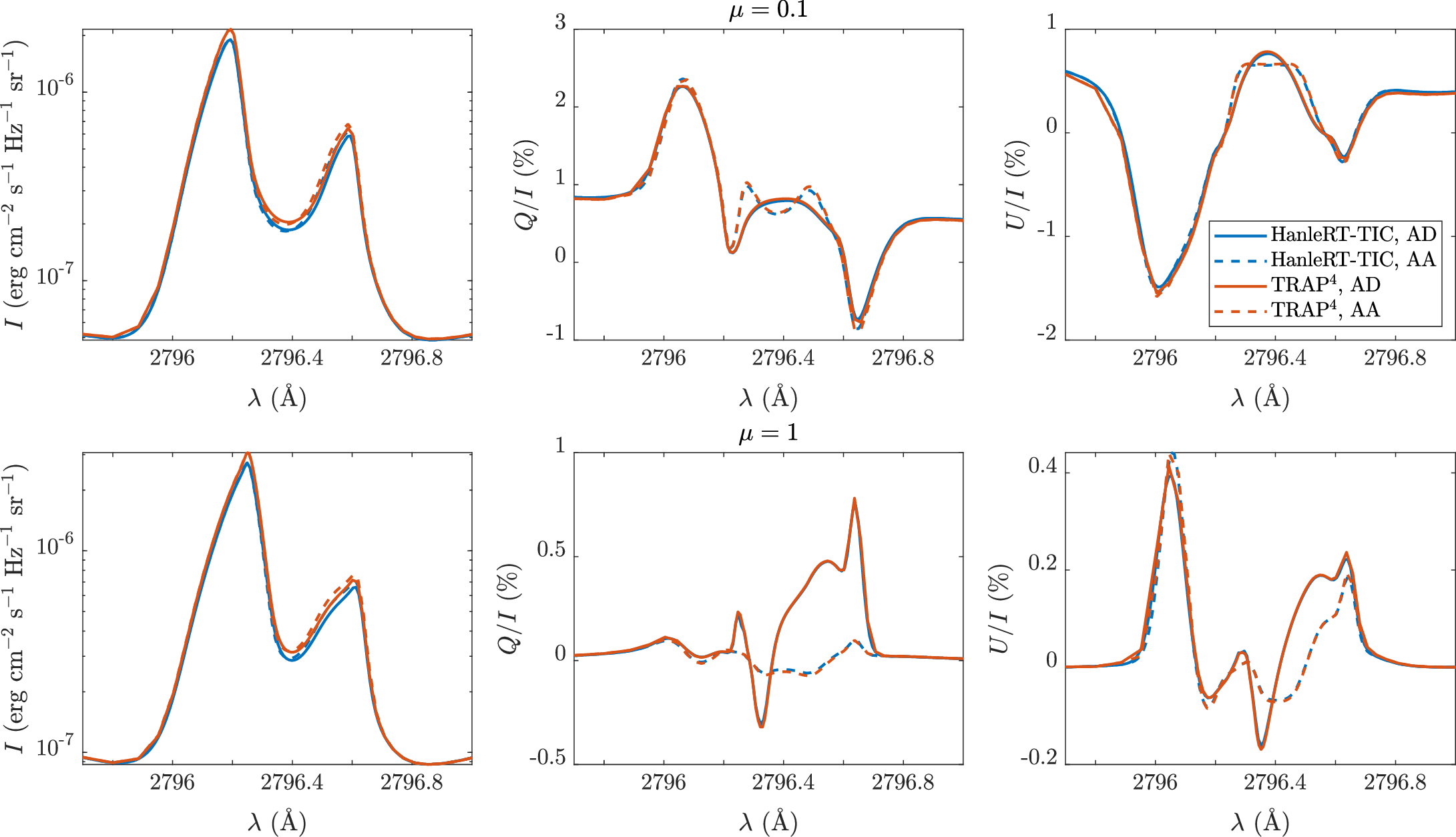}
    \caption{Intensity (left column), $Q/I$ (middle column), and $U/I$ (right column) as a function of vacuum wavelength for the Mg~{\sc ii} k line obtained with PRD--AA (dashed lines) and PRD--AD (solid lines) calculations in a dynamic magnetic atmosphere (see text) for the two LOSs with $\mu=0.1$ (first row) and $\mu=1$ (second row). The blue and red curves correspond to calculations carried out with the HanleRT-TIC and the TRAP$^4$ code, respectively. The reference direction for positive Stokes $Q$ is the parallel to the limb.}
    \label{fig:benchmark}
\end{figure*}

The results are shown in Fig.~\ref{fig:benchmark} for the k line at the two LOSs $\mu=0.1$ (first row) and $\mu=1$ (second row), with the latter denoting a forward-scattering geometry~\citep[see, e.g.][]{belluzzi2024}.
Here, we only focus on the k line, as it is the spectral region showing the largest discrepancies between the results of the two codes.
Overall, we found very satisfying agreement between HanleRT-TIC (blue curves) and TRAP$^4$ (red curves), 
in particular for the fractional polarisation signals (middle and right columns).
On the other hand, 
relative differences of the order of $5-10\%$
are visible in $I$ near the k line core.
This discrepancy is probably related to the different normalisation of the emissivity and formulation of the redistribution matrices used in the two codes.

The comparison between the PRD--AA (dashed lines) and PRD--AD~(solid lines) 
results reveals once more the importance of using an AD description of scattering processes when modelling
the Mg~{\sc ii} h\&k doublet in a dynamic and magnetic atmosphere, as discussed in great detail in~\citet{delpinoaleman2024}.
Indeed, it is particularly important in order to accurately model the 
forward-scattering Hanle effect~\citep[see $Q/I$ and $U/I$ at $\mu=1$, as well as the discussion in][]{belluzzi2024}.

\subsection{Validation of the two-step approach}
\label{sec:validation_solution}
It is important to note that, in agreement with our assumption that polarisation has a negligible impact on the ground-level population of Mg~{\sc ii}, the input parameters of the TRAP$^4$ code
were obtained from HanleRT-TIC PRD calculations for a static, unmagnetic atmosphere and neglecting polarisation, as detailed in Sect.~\ref{sec:input_parameters}. 
On the other hand, the HanleRT-TIC results in Fig.~\ref{fig:benchmark} (blue lines) were obtained calculating the ground-level population in a self-consistent way, accounting for the impact of polarisation, magnetic fields, and velocities.
Remarkably, we found relative differences in the ground-level populations between the two cases that are below $0.2\%$.
A detailed investigation also revealed that these extremely small differences do not play any role in the Stokes profiles computed with TRAP$^4$.
This fact, in combination with the excellent agreement between the HanleRT-TIC and TRAP$^4$ results of Sect.~\ref{sec:validation_HanleRT}, is thus a validation of 
the working assumptions discussed in Sect.~\ref{sec:linearisation} and of our solution strategy.

\subsection{Impact of PRD--AD on the Ly-$\alpha$ line}
\label{sec:AA_AD_HI}
As an application of broad physical interest, we also investigated the suitability 
of using the AA approximation when modelling wavelength-integrated scattering polarisation signals 
of the H~{\sc i} Ly-$\alpha$ line, as done by~\citet{alsina_ballester2023}.
\begin{figure*}[ht!]
    \centering
    \includegraphics[width=0.95\textwidth]{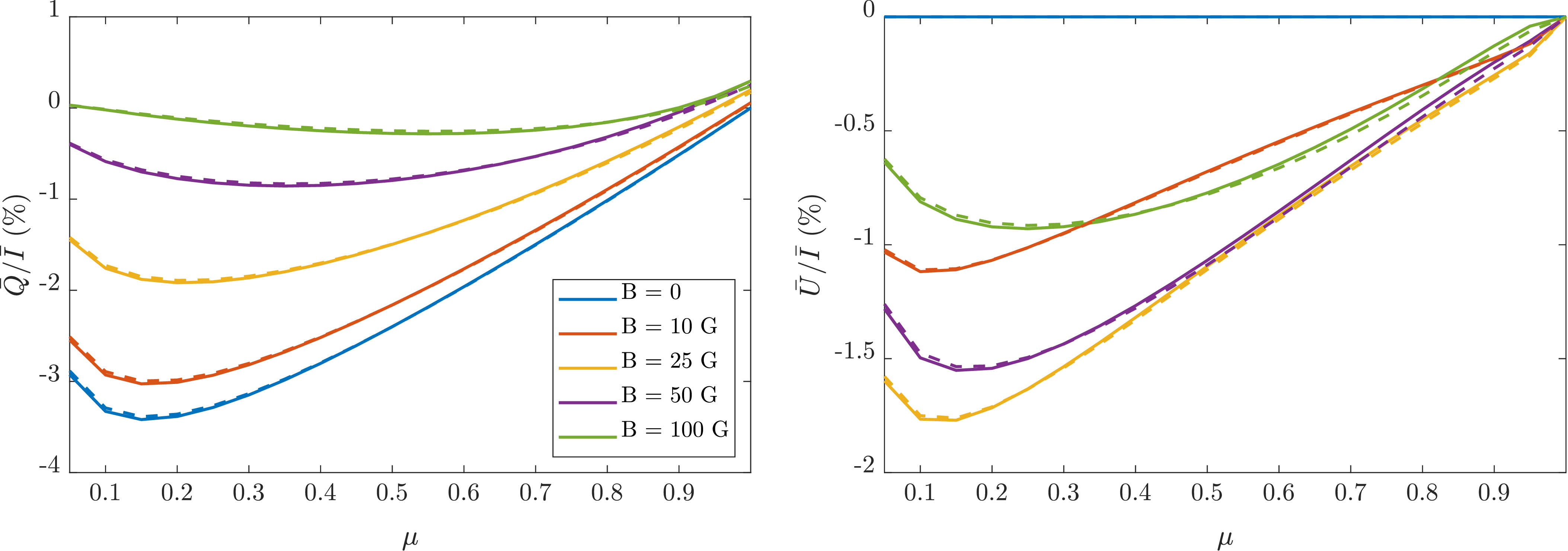}
    \caption{Centre-to-limb variation of $\overline{Q}/\overline{I}$ (left panel) and $\overline{U}/\overline{I}$ (right panel) as function of $\mu=\cos(\theta)$. The results are obtained for $\chi=0$ and considering height-independent horizontal ($\theta_B=\pi/2$ and $\chi_B=0$) magnetic fields of different strengths (see the legend).
 Dashed and solid lines correspond to PRD--AA and PRD--AD calculations, respectively.}
    \label{fig:centre_to_limb}
\end{figure*}
To this aim, we computed the emergent Stokes profiles of the H~{\sc i} Ly-$\alpha$ line
obtained with a static magnetic atmosphere, with height-independent horizontal magnetic fields
of different strengths $B=0,\ 10,\ 25,\ 50,$ and $100\ \mathrm{G}$, both within the PRD--AA and PRD--AD descriptions.
Then, following~\citet{alsina_ballester2023}, synthetic narrow-band signals are obtained by weighting by a Gaussian 
the numerical emergent Stokes profiles and integrating the results over wavelength.
Namely,
\begin{equation*}
\overline{X}(\mu)=\frac{1}{\sqrt{2\pi}\sigma_\lambda}
\int_{\lambda_0-\Delta\lambda}^{\lambda_0+\Delta\lambda}\exp\left(-\frac{(\lambda-\lambda_0)^2}{2\sigma_\lambda^2}\right)X(z_{\max},\mu,\lambda)\,{\rm d}\lambda,
\end{equation*}
where $X = I, Q,$ and $U$, $\lambda_0 = 1215.67$~\AA~
is the H~{\sc{i}} Ly-$\alpha$ line-centre wavelength in vacuum,
and $\sigma_\lambda$
is the standard deviation of the
Gaussian weighting function, selected so that it
corresponds to a full width at half maximum of $2\sqrt{2\log(2)}\sigma_\lambda=35\ \AA$.
For the present study, we considered as integration limit $\Delta\lambda=30\ \AA$~\citep[see][for a discussion of the impact of $\Delta\lambda$ on the results]{alsina_ballester2023}.

The resulting centre-to-limb profiles are displayed in Fig.~\ref{fig:centre_to_limb}, 
with dashed and solid lines for PRD--AA and PRD--AD calculations, respectively.
Overall, there is excellent agreement between PRD--AA and PRD--AD results.
Minor discrepancies are found, in particular for $\overline{U}/\overline{I}$ near disk centre ($\mu\sim1$), 
but their amplitude is small.
We also repeated the computations for different values of $\Delta\lambda$ and with the atmospheric models A, F, and P of~\citet{fontenla1993}, finding negligible differences between PRD--AA and PRD--AD results.

\section{Conclusions}
\label{sec:conclusions}
In the present paper, we illustrate an accurate and effective solution strategy 
to model the Stokes profiles of strong resonance lines originating in the solar atmosphere, 
taking into account scattering polarisation with PRD effects in their general AD formulation, $J$-state interference,
and the impact of arbitrary magnetic fields and bulk velocities.
The formalism we introduce can be applied both to two-level and two-term atoms, also allowing the inclusion of atomic hyperfine structure.
Moreover, although the formalism is here applied to 1D plane-parallel atmospheric models,
it can be easily generalised to 3D geometries \citep[see][]{benedusi2023}.

After introducing the new formalism, a suitable physical approximation,
thanks to which the problem becomes linear in the radiation field $\bm{I}$, is discussed.
The approximation consists in 
assuming that polarisation has a negligible impact on the population of the lower level/term of resonance lines, 
so that it can be retrieved from independent unpolarised calculations and considered as an input parameter of the problem.
Together with a suitable algebraic formulation of the problem,
this allows us to exploit the innovative physics-based preconditioners of~\citet{janett2024} and 
efficiently solve the problem with matrix-free preconditioned algorithms.
Since the computational time necessary to apply the physics-based preconditioner to the full problem 
can be a large fraction of the full time-to-solution, we also introduce an additional block-diagonal preconditioner, 
successfully used to speed up the calculations.

To accelerate the calculations in the presence of bulk velocities, 
we compute the scattering emissivity in the comoving frame and then interpolate it to the observer's frame.
This strategy drastically reduces the required number of evaluations of the redistribution functions.
Moreover, we present a careful choice of the angular and spectral quadrature nodes that, in combination with a proper normalisation of the scattering emissivity,
allows for a solution of the problem that is at the same time fast and accurate.

The presented solution strategy is applied to the synthesis of the Stokes profiles of the Mg~{\sc ii} h\&k doublet
and the H~{\sc i} Ly-$\alpha$ line.
In particular, considering the 1D FAL-C atmospheric model, 
we performed a solution verification of the emerging synthetic profiles,
showing that, for the considered numerical parameters, the results are solid and 
the numerical errors considerably small.
A detailed benchmark of TRAP$^4$ with HanleRT-TIC results also displays good agreement between the two codes.
This is a remarkable validation of the linearisation strategy outlined in Sect.~\ref{sec:linearisation}.

Finally, wavelength-integrated polarisation profiles of the H~{\sc i} Ly-$\alpha$ line 
show excellent agreement when comparing PRD--AA and PRD--AD calculations.
The very minor discrepancies between the two computations
suggest that a complete PRD--AD description of scattering processes is not strictly necessary 
to accurately model the impact of magnetic fields with $B\leq100\ \mathrm{G}$ on these signals, thus validating the work of \citet{alsina_ballester2023}.
However, this comparison was limited to 1D plane-parallel atmospheric models and it will be important to verify whether this conclusion also remains valid in comprehensive 3D models,
for example by implementing the presented solution strategy in
the novel 3D numerical framework of~\citet{benedusi2023}.

\begin{acknowledgements}
The financial support by the Swiss National Science Foundation (SNSF) through grant CRSII5\_180238 is gratefully acknowledged.
T.P.A., E.A.B., and J.T.B. acknowledge support from the Agencia Estatal de Investigación
del Ministerio de Ciencia, Innovación y Universidades (MCIU/AEI) under grant
``Polarimetric Inference of Magnetic Fields'' and the European Regional Development
Fund (ERDF) with reference PID2022-136563NB-I00/10.13039/501100011033.
E.A.B. acknowledges support from the European Research Council through the Synergy grant No. 810218 (“The Whole Sun” ERC-2018-SyG).
T.P.A.'s participation in the publication is part of the Project RYC2021-034006-I,
funded by MICIN/AEI/10.13039/501100011033, and the European Union ``NextGenerationEU''/RTRP.
\end{acknowledgements}

\bibliographystyle{aa}
\bibliography{bibfile}

\appendix

\section{Continuum processes}
\label{app:continuum}
Under typical conditions of the solar atmosphere, 
it is a good approximation to assume that continuum processes do not contribute to dispersion and dichroism, that is, to the off-diagonal elements of $K$ \citep[e.g.][]{landi_deglinnocenti+landolfi2004}.
Additionally, the continuum contribution to the diagonal elements of~$K$ (continuum opacity) is given by $\eta_1^c(\rr,\nu)=\kappa^c(\rr,\nu)+\sigma^c(\rr,\nu)$, where $\kappa^c$ and $\sigma^c$ are the isotropic extinction coefficients for true absorption and scattering, respectively, and the superscript $c$ denotes continuum contributions.

Concerning the continuum emissivity, we consider both thermal and scattering processes,
for which we make the assumption of coherent scattering in the observer's frame.
This implies that the continuum contribution to the emissivity can be written as
\begin{align*}
\pmb{\varepsilon}^c(\rr,\vec{\Omega},\nu) = & \ \pmb{\varepsilon}^{c,\text{th}}(\rr,\nu) + \sigma^c(\rr,\nu)\!\sum_{K,Q}\!(-1)^Q \pmb{\mathcal{T}}_Q^K(\vec{\Omega})\\
& \times \!\oint\! \pmb{\mathcal{T}}_{-Q}^K(\vec{\Omega}')\cdot\bm{I}(\rr,\vec{\Omega}',\nu)\frac{{\rm d} \vec{\Omega}'}{4 \pi},
\end{align*}
where $\pmb{\varepsilon}^{c,\text{th}}$ is the continuum thermal emissivity (which is isotropic and only contributes to Stokes $I$) and $\pmb{\mathcal{T}}$ is a vector whose four components are the polarisation tensors~\citep[see Section 5.11 of][]{landi_deglinnocenti+landolfi2004} in the four Stokes parameters.
We note that $\kappa^c$, $\sigma^c$, and $\pmb{\varepsilon}^{c,\text{th}}$ are currently evaluated with publicly available routines~\citep[e.g.][]{uitenbroek2001,li22}.

\section{$A_Q^{KK'}$, $B_Q^{KK'}$, and branching ratios}
\label{app:AQKKp}

The quantity $A_Q^{KK'}$ appearing in Eq.~\eqref{eq:RQIIKKp} is defined as
\begin{equation*}
A_Q^{KK'}(\rr,k_u,k_u',k_\ell,k_\ell')=\frac{\alpha_{k_u,k_u'}(\rr)}{2\pi\Delta\nu_D(\rr)^2}\mathcal{A}_Q^{KK'}(k_u,k_u',k_\ell,k_\ell'),
\end{equation*}
where $\alpha_{k_u,k_u'}$ represents the branching ratio for scattering processes described by $\mathcal{R}^{{\scriptscriptstyle \rm II}}$ (it contains both the branching ratio 
between thermal and scattering processes and the one between $\mathcal{R}^{{\scriptscriptstyle \rm II}}$ and $\mathcal{R}^{{\scriptscriptstyle \rm III}}$)
and $\mathcal{A}_Q^{KK'}$ is a quantity depending on a series
of quantum numbers characterising the involved atomic states.

According to the works of \citet{bommier1997a,bommier1997b,bommier2017}, the branching ratio for $\mathcal{R}^{{\scriptscriptstyle \rm II}}$ is given by 
\begin{equation*}
\alpha_{k_u,k_u'}(\rr)=\frac{\Gamma_\mathrm{R}}{\Gamma_\mathrm{R}+\Gamma_\mathrm{I}(\rr)+\Gamma_\mathrm{E}(\rr)+2\pi i\nu_{k_u,k_u'}(\rr)}.
\end{equation*}
With this definition, and taking
$\mathcal{A}_Q^{KK'}(k_u,k_u',k_\ell,k_\ell') = \mathcal{C}_{QKK'M_uM_u'M_\ell M_\ell'pp'p''p'''}$ \citep[see Eq.~(12) of][]{bommier1997b},
one recovers an expression for $\mathcal R_Q^{{\scriptscriptstyle \rm II},KK'}$ fully analogous to Eq.~(23) of~\citet{alsinaballester2017} for the two-level atom. 
Similarly, by considering 
$\mathcal{A}_Q^{KK'}(k_u,k_u',k_\ell,k_\ell')=\mathcal{A}_Q^{KK'}(j_uM_u,j_u'M_u',j_\ell M_\ell,j_\ell'M_\ell')$ \citep[see Eq.~(C.19) of][]{alsinaballester2022},
one recovers the redistribution functions in Eq.~(10) of~\citet{alsinaballester2022} for a two-term atom.\footnote{The only difference being that, here, we also account for plasma bulk velocities, which introduce a reduced Doppler shift frequency $u_\mathrm{b}$.}

The quantity $B_Q^{KK'}$ appearing in the definition of $\mathcal R_Q^{{\scriptscriptstyle \rm III},KK'}$, Eq.~\eqref{eq:RIIIQKKp}, is defined as
\begin{equation*}
B_Q^{KK'}(\rr,k_u,k_u',k_\ell,k_\ell')=\frac{\beta_{k_u,k_u'}(\rr)-\alpha_{k_u,k_u'}(\rr)}{4\pi\Delta\nu_D(\rr)^2}\mathcal{A}_Q^{KK'}(k_u,k_u',k_\ell,k_\ell'),
\end{equation*}
where the term $\beta_{k_u,k_u'}-\alpha_{k_u,k_u'}$ is the branching ratio for $\mathcal{R}^{{\scriptscriptstyle \rm III}}$, with
\begin{equation*}
\beta_{k_u,k_u'}(\rr)=\frac{\Gamma_\mathrm{R}}{\Gamma_\mathrm{R} + \Gamma_\mathrm{I}(\rr) + 2\pi i\nu_{k_u,k_u'}(\rr)}.
\end{equation*}
Here, we assumed that the depolarising effect of elastic collisions is negligible. 
Notably, this is a good approximation when modelling the polarisation of strong resonance lines that form in the low-density plasma of the solar chromosphere \citep{alsinaballester2021}.
We note that the expression we provided for $\beta_{k_u,k_u'}$ holds both for a two-level atom \citep{bommier1997b} and for a two-term atom \citep{bommier2017} and include the term $1-\epsilon$.

If depolarising collisions are not negligible, the redistribution matrix formalism cannot be used for a two-term atomic model, as the SE equations do not have a closed analytic solution \citep[unless magnetic fields and collisional transitions between different $J$-levels of the same term are neglected; see][]{bommier2017}.
However, the redistribution matrix can be derived also accounting for depolarising collisions in the case of a two-level atom~\citep[see][]{bommier1997b}, the only difference being in the $\beta_{k_u,k_u'}$ factor, which takes the form
\begin{equation*}
\beta^{(K)}_{k_u,k_u'}(\rr)=\frac{\Gamma_\mathrm{R}}{\Gamma_\mathrm{R} + \Gamma_\mathrm{I}(\rr) + D^{(K)}(\rr) + 2\pi i\nu_{k_u,k_u'}(\rr)}, \,
\end{equation*}
with $D^{(K)}$ the depolarisation rate due to elastic collisions. 
With these branching ratios, one recovers the expressions for $\mathcal R^{\scriptscriptstyle \rm III}$ in Eq.~(20) of~\citet{alsinaballester2017} and in Eq.~(11) of~\citet{alsinaballester2022} for two-level and two-term atoms, respectively.

\section{Spectral quadrature grid for $\mathcal R^{\scriptscriptstyle \rm II}$}
\label{app:spectral_quadratrue}

Here, we detail the spectral quadrature grids $\{u_{n'}\}_{n'=1}^{N_{\nu'}}$ used to numerically evaluate Eq.~\eqref{eq:int_Ftilde}. Recalling that $u_n$ denotes a discrete point of the reduced frequency grid, these are defined as follows:

\textbf{i.} if $\Theta=0$, Eq.~\eqref{eq:int_Ftilde} is integrated analytically;

\textbf{ii.} if $\Theta\geq\pi/8$ or $\min_i|u_n+u_i^*|\geq u^\mathrm{t}$, case~(d) in Sect.~\ref{sec:spectral_quadrature}, we use a Gauss-Hermite quadrature with 65 nodes;

\textbf{iii.} if $\Theta>\pi/8$ and $\min_i|u_n+u_i^*|\leq2$, we are in the line-core regime, case~(e) in Sect.~\ref{sec:spectral_quadrature}, and we distinguish between the three different cases 
\begin{itemize}
\item $\pi/8\leq\Theta\leq\pi/2$, for which a single GL quadrature of 101 nodes over the interval $[u^\mathrm{M}-\Delta^1,u^\mathrm{M}+\Delta^1]$ is used,
\item $\pi/2<\Theta\leq15\pi/16$, for which three contiguous GL quadratures of 55, 81, and 55 nodes over the intervals $[u-\Delta^1,u^\mathrm{M}-\Delta^2]$, $[u^\mathrm{M}-\Delta^2,u^\mathrm{M}+\Delta^2]$, and $[u^\mathrm{M}+\Delta^2,u+\Delta^1]$, respectively, are used, and
\item $15\pi/16<\Theta<\pi$, for which four contiguous GL quadratures of 65, 101, 101, and 65 nodes over the intervals $[u-\Delta^1,u^\mathrm{M}-\Delta^3]$, $[u^\mathrm{M}-\Delta^3,u^\mathrm{M}]$, $[u^\mathrm{M},u^\mathrm{M}+\Delta^3]$, and $[u^\mathrm{M}+\Delta^3,u+\Delta^1]$, respectively, are used;
\end{itemize}

\textbf{iv.} if $\Theta>\pi/8$ and $2<\min_i|u_n+u_i^*|<u^\mathrm{t}$, we are in the near-wing regime, case (f) in Sect.~\ref{sec:spectral_quadrature}, and we combine three GL contiguous quadratures of 55, 95, 121 nodes (121, 95, 55 if $u_n+u_i^*<0$) over the intervals $[L^\mathrm{nw},u^\mathrm{M}-\Delta^4]$, $[u^\mathrm{M}-\Delta^4,u^\mathrm{M}+\Delta^4]$, and $[u^\mathrm{M}+\Delta^4,R^\mathrm{nw}]$, respectively.

Here, we introduced the threshold $u^\mathrm{t}=\max(5.8,2.5\,\Theta)$ and we defined
$u^\mathrm{M}=(u_n+u_i^*)\cos(\Theta)-u_i^*$, 
$\Delta^1=11\sin(\Theta/2)$, 
$\Delta^2=6(0.1+a+\cos(\Theta/2))/(a+1)$, 
$\Delta^3=6(0.05+a+\cos(\Theta/2))/(a+1)$, 
$\Delta^4=6\cos(\Theta/2)+0.2$,
$L^\mathrm{nw}=\min(u^\mathrm{M}-\Delta^4-1,\tilde{u}-\Delta^1)$, and
$R^\mathrm{nw}=\max(u^\mathrm{M}+\Delta^4+1,\tilde{u}+\Delta^1)$,
where $u_i^*$ is the cluster centre closest to $-u_n$ and $\tilde{u}=u_\mathrm{Re}$ and $\tilde{u}=u_\mathrm{Im}$ for the real and imaginary parts of Eq.~\eqref{eq:int_Ftilde}, respectively, being
\begin{align*}
&u_{\rm Re}=\mathrm{sign}(u_n+u_i^*)\sqrt{(u_n+u_i^*)^2 - 4\sin^2(\Theta/2)},\\
&u_{\rm Im}=\mathrm{sign}(u_n+u_i^*)\sqrt{(u_n+u_i^*)^2 - 2\sin^2(\Theta/2)}.
\end{align*}

Concerning the AA approximation,
the integral over $\Theta$ in Eq.~\eqref{eq:R_AA}
smooths out the sharp peaks and sign reversals
of $\mathcal{\bar R}^{\scriptscriptstyle \rm II}$ near $\Theta=0$ and $\Theta=\pi$, 
thus simplifying the numerical integration
of $\mathcal{\bar R}^{\scriptscriptstyle \rm II-AA}\cdot I$ over frequency
with respect to the general AD case.
The integral in Eq.~\eqref{eq:R_AA} also drastically reduces the degree of freedom
of the redistribution function, removing its dependence on~$\Theta$ and making it possible to assemble
$\mathcal{\bar R}^{{\scriptscriptstyle \rm II-AA}}$ only once, before starting the iterative process,
and simply storing it in memory. 
For the selection of $\{u_{n'}\}_{n'=1}^{N_{\nu'}}$, we thus opted for a compromise between minimum number of quadrature nodes, simplicity, and accuracy. Specifically, we used two GL quadratures of 142 nodes each in the two intervals
$[u_n-7.8,u_n]$ and $[u_n,u_n+7.8]$. 
For the integral over $\Theta$, we opted for one GL quadrature of 25 nodes in the interval~$[0,\pi]$.

\section{Additional optimisation strategies}
\label{app:strategy}

In this appendix, we neglect the dependence of any quantity on~$\rr$ for ease of notation.
First, to further optimise the code, we note that 
$\mathcal{R}^{{\scriptscriptstyle \rm II},KK'}_{-Q}\mathcal{P}^{KK'}_{-Q}+\mathcal{R}^{{\scriptscriptstyle \rm II},KK'}_Q\mathcal{P}^{KK'}_{Q}=2\mathrm{Re}(\mathcal{R}^{{\scriptscriptstyle \rm II},KK'}_{|Q|}\mathcal{P}^{KK'}_{|Q|})$.
This is exploited to avoid evaluating $\mathcal{R}^{{\scriptscriptstyle \rm II},KK'}_Q$ for negative $Q$.
Additionally, to reduce the number of evaluations of the Faddeeva function, of integrals over frequency, and of interpolations, we define the function
\begin{align}
\vec{\mathcal W}_{k_u,k_\ell}(\vec{\Omega}',\vec{\Omega},u_n)=&\Delta\nu_D\int F\left(\Theta,u'+u_{k_u,k_\ell},u_n+u_{k_u,k_\ell}\right)\nonumber\\
&\times\vec{I}\left(\vec{\Omega}',u'-u_\mathrm{b}(\vec{\Omega}')\right)\mathrm{d}u'.\label{eq:Wkukl}
\end{align}
Using this definition, we rewrite the $R^{{\scriptscriptstyle \rm II}}$ contribution to Eq.~\eqref{eq:eps_sc_l} as
\begin{align}\label{eq:eps_sc_II_exp}
\pmb{\varepsilon}^{\ell,{\rm sc,II}}(\vec{\Omega},\nu_n) \!=&\,
        k \sum_{K,K',Q}\sum_{k_uk_u'}\sum_{k_\ell k_\ell'}
        A_{|Q|}^{KK'}(k_u,k_u',k_\ell,k_\ell')\nonumber\\
&\times
        \oint\!\frac{{\rm d} \vec{\Omega}'}{4 \pi}\mathcal{P}^{KK'}_{|Q|}(\vec{\Omega}',\vec{\Omega})\nonumber\\
&\times\left[\vec{\mathcal W}_{k_u',k_\ell}\left(\vec{\Omega}',\vec{\Omega},u_n+u_\mathrm{b}(\vec{\Omega})+u_{k_\ell,k_\ell'}\right)\right.\nonumber\\
&+\left.\vec{\mathcal W}_{k_u,k_\ell}\left(\vec{\Omega}',\vec{\Omega},u_n+u_\mathrm{b}(\vec{\Omega})+u_{k_\ell,k_\ell'}\right)^*\right].
\end{align}
In practice, we never fully assemble the redistribution function $R_Q^{{\scriptscriptstyle \rm II},KK'}$ in the code.
Instead, we first evaluate $F\left(\Theta,u_{n'}+u_{k_u,k_\ell},u_n+u_{k_u,k_\ell}\right)$ and $\vec{I}\left(\vec{\Omega}',u_{n'}-u_\mathrm{b}(\vec{\Omega}')\right)$ separately.
Then, we perform the quadrature in frequency of the product of the two quantities to get $\vec{\mathcal W}_{k_u,k_\ell}$. 
Before computing the sum over $k_\ell$ and $k_\ell'$, one has to interpolate $\vec{\mathcal W}_{k_u,k_\ell}$ from $u_n$ to $u_n+u_\mathrm{b}(\vec{\Omega})+u_{k_\ell,k_\ell'}$.
Finally, $\pmb{\varepsilon}^{\ell,{\rm sc,II}}$ is evaluated according to Eq.~\eqref{eq:eps_sc_II_exp}.
This strategy allows us to evaluate $\pmb{\varepsilon}^{\ell,{\rm sc,II}}$ by computing the Faddeeva function and interpolating $\vec{I}$ as few times as possible, while still ensuring
an accurate result.

\section{Normalisation of line emissivity}
\label{app:normalisation}

In LTE, in the presence of a magnetic field, Kirchhoff's law~\citep{kirchhoff1860} can be generalised by
accounting for the Zeeman effect and for polarisation~\citep[see Sect.~9.1 of][]{landi_deglinnocenti+landolfi2004}, that is,
\begin{equation*}
    \pmb{\varepsilon}(\rr,\vec{\Omega},\nu) = W_T(\rr) \vec{\eta}(\rr,\vec{\Omega},\nu),
\end{equation*}
where the Planck function in the Wien limit ($W_T$) is used because stimulated emission is neglected.
The dependence of $W_T$ on wavelength is also neglected.
Moreover, considering that
the branching ratio of scattering contributions to the emission vector is $1-\epsilon$, we can write
\begin{equation}\label{analytical_eps_sc}
    \pmb{\varepsilon}^{\rm sc}(\rr,\vec{\Omega},\nu) = \left[1-\epsilon(\rr)\right]W_T(\rr) \vec{\eta}(\rr,\vec{\Omega},\nu).
\end{equation}
This relation must hold not only when considering 
the sum of line and continuum contributions, 
but also when we consider the two contributions individually.
In TRAP$^4$, we account for numerical deviations
from Kirchhoff's law in the line emission contribution
in the following way:
\begin{enumerate}
    \item we calculate the reference for the scattering contribution to line emissivity, $\pmb{\varepsilon}^{\ell,\rm sc, ref}$,
    with~Eq.~\eqref{analytical_eps_sc};
    \item we assume LTE by setting a Planckian incident radiation field (i.e. unpolarised, flat, and isotropic), namely,
$\bm{I}(\rr,{\bf\Omega},\nu)= (W_T(\rr), 0, 0, 0)$,
    and we accordingly calculate the line scattering contribution to emissivity, $\pmb{\varepsilon}^{\ell,\rm sc, num}$, with the code;
    \item we calculate the normalisation factor $\bm{f}\in\mathbb R^N$, which is the ratio between Kirchhoff (calculated at 1.)
    and numerical (calculated at 2.) scattering contribution to line emissivity, namely
    $$\bm{f}(\rr,\vec{\Omega},\nu)=\frac{\pmb{\varepsilon}^{\ell,\rm sc, ref}(\rr,\vec{\Omega},\nu)}{\pmb{\varepsilon}^{\ell,\rm sc,num}(\rr,\vec{\Omega},\nu)};$$
    \item we multiply $\pmb{\varepsilon}^{\ell,\rm sc}$, calculated via Eq.~\eqref{eq:eps_sc_l}, by the normalisation factor $\bm{f}$ when solving iteratively the linear system given by Eq.~\eqref{eq:linear_system}.
\end{enumerate}
Since the normalisation factor only depends on the input parameters of the problem (i.e. it does not depend on the iteration), it can be computed a priori and then applied at each iteration.
We note that, for practical applications, we only normalised the intensity component of the emission vector.

\section{Block-diagonal preconditioner}
\label{app:Block_jacobi}
In order to apply the physics-based preconditioner $P$ to the RT problem given by Eq.~\eqref{eq:linear_system},
we need to routinely solve a size-$N$ linear system of the form $P\bm{x}=\bm{y}$.
This task may prove computationally expensive even when $P$ corresponds to the simplified
PRD--AA setting (see Sect.~\ref{sec:preconditioning}).
To overcome this issue, we applied a block-diagonal preconditioner $\hat P$, similar to that proposed by~\citet[][]{alsinaballester2017}, to the linear system $P\bm{x}=\bm{y}$, that is, we solved the equivalent problem
$$\hat P^{-1}P\bm{x}=\hat P^{-1}\bm{y},$$
with
$\hat{P}=\mathrm{diag}\left(\hat P_1,...,\hat P_{N_r}\right)$
an invertible block-diagonal matrix and $\hat P_i\in \mathbb R^{4N_\Omega N_\nu\times 4N_\Omega N_\nu}$ the spatially local preconditioner
operating at the discrete spatial point $\rr_i\in D$ (with $i=1,\ldots,N_r$).

To conceive $\hat{P}$, we write the spatial components of the scattering emissivity in the observer's frame,
$\pmb{\varepsilon}_i^{\rm sc}\in\mathbb R^{4N_\Omega N_\nu}$ ($i=1,\ldots,N_r$),
in the compact matrix form
\begin{equation*}
\pmb{\varepsilon}_i^{\rm sc}=\Sigma_i^*\bm{I}_i,
\end{equation*}
where we introduced the local Stokes parameters in the observer's frame $\bm{I}_i\in\mathbb R^{4N_\Omega N_\nu}$
and the local linear scattering operator in the observer's frame $\Sigma_i^*\in\mathbb R^{4N_\Omega N_\nu\times 4N_\Omega N_\nu}$ (i.e. $\Sigma_i^*$ is the local block of the approximated scattering operator $\Sigma^*$ used to build $P$, see Sects.~\ref{sec:algebraic} and \ref{sec:preconditioning}).
Then, we rewrite the transfer equation given by Eq.~\eqref{eq:matrix_form_1} as
\begin{equation*}
\bm{I}_i=\sum_{j=1}^{N_r}\Lambda_{ij}(\Sigma_j^*\bm{I}_j+\pmb{\varepsilon}_j^\mathrm{th})+\bm{t}_i,
\end{equation*}
where the diagonal matrices $\Lambda_{ij}\in\mathbb R^{4N_\Omega N_\nu\times 4N_\Omega N_\nu}$ encode the formal solution
(namely, the transfer from $\rr_j$ to $\rr_i$, not to be confused with $\tilde\Lambda_{ij}$ introduced in Sect.~\ref{sec:algebraic})
and the vectors $\pmb{\varepsilon}_i^\mathrm{th}\in\mathbb R^{4N_\Omega N_\nu}$ and $\bm{t}_i\in\mathbb R^{4N_\Omega N_\nu}$ 
represent the local thermal emissivity and the radiation transmitted 
from the boundaries to~$\rr_i$, respectively.
Within this formalism, the block-Jacobi method can be simply written as
\begin{equation}\label{eq:Block_Jacobi}
\bm{I}_i^{n+1}=\sum_{j=1}^{N_r}\Lambda_{ij}(\Sigma_j^*\bm{I}_j^n+\pmb{\varepsilon}_j^\mathrm{th})+\bm{t}_i+\Lambda_{ii}\Sigma_i^*(\bm{I}_i^{n+1}-\bm{I}_i^n),
\end{equation}
where the superscripts $n+1$ and $n$ denote the values of $\bm{I}_i$ at two successive iterations. 
Equation~\eqref{eq:Block_Jacobi} can readily be recast into
\begin{equation}\label{eq:Delta_I}
(I\hspace{-0.1em}d^{4N_\Omega N_\nu}-\Lambda_{ii}\Sigma_i^*)(\bm{I}_i^{n+1}-\bm{I}_i^n)=\bm{R}_i^n,
\end{equation}
thus identifying the local blocks of the block-Jacobi preconditioner as $\hat P_i=I\hspace{-0.1em}d^{4N_\Omega N_\nu}-\Lambda_{ii}\Sigma_i^*\in\mathbb R^{4N_\Omega N_\nu\times 4N_\Omega N_\nu}$. Here, $I\hspace{-0.1em}d^{4N_\Omega N_\nu}\in\mathbb R^{4N_\Omega N_\nu\times 4N_\Omega N_\nu}$ is the identity matrix and the vector $\bm{R}_i^n=\sum_{j=1}^{N_r}\Lambda_{ij}(\Sigma_j^*\bm{I}_j^n+\pmb{\varepsilon}_j^\mathrm{th})+\bm{t}_i-\bm{I}_i^n\in\mathbb R^{4N_\Omega N_\nu}$ represents the local residual of the RT problem encoded in $P$ at iteration $n$ and at~$\rr_i$.

Unfortunately, the application of $\hat P^{-1}$ would require inverting, and possibly storing, $N_r$ matrices of size $4N_\Omega N_\nu\times 4N_\Omega N_\nu$. While this can be done for small 1D plane-parallel problems, it would be computationally unfeasible for larger spatial domains. 
To overcome this issue, we proceed as follows.
First, we further simplify the problem by considering only the unmagnetic and unpolarised setting.
Within this approximation, we only consider the intensity component of all vectors and matrices introduced above (i.e. vector dimensions reduce by a factor of four and matrix dimensions reduce by a factor of $4\times4$).
Moreover, in the unmagnetic and unpolarised case, the scattering phase matrix $\mathcal{P}^{KK'}_{Q}$ is independent of the spatial location and reduces to a complex scalar of value $(-1)^Q\mathcal{T}_{Q,1}^{K'}(\vec{\Omega})\mathcal{T}_{-Q,1}^K(\vec{\Omega}')$, where $\mathcal{T}_{Q,i}^{K}$ ($i=1,...,4$) are the geometrical tensors in the observer's frame~\citep[see Sect. 5.11 of][]{landi_deglinnocenti+landolfi2004}.
Second, we note that the AA redistribution functions in the operator $\Sigma^*$, 
when evaluated in the comoving frame~(see Sect.~\ref{sec:bulkvel}),
are independent of the scattering angle $\Theta$.
Thus, observing that in the unmagnetic case only the 
components of the redistribution function with $K'=K$ are non-zero, we rewrite the scattering emissivity in the comoving frame, Eq.~\eqref{eq:eps_sc_bar}, for the unmagnetic and unpolarised setting as
\begin{align*}
\bar{\pmb{\varepsilon}}^{{\rm sc}}(\rr,\vec{\Omega},\nu) = &\ k(\rr) 
        \!\!\sum_{K,Q}\!(-1)^Q\mathcal{T}_{Q,1}^{K}(\vec{\Omega})\\
        &\times \!\int\!
	[\mathcal{\bar R}^*]^{KK}_Q(\rr,\nu',\nu) 
        \bar{J}_{-Q}^K(\rr,\nu'){\rm d} \nu',
\end{align*}
where we introduced the radiation field tensor for the unpolarised setting in the comoving frame $\bar{J}_Q^K(\rr,\nu)\!=\!\oint \mathcal{T}_{Q,1}^K(\vec{\Omega})\bar{I}(\rr,\vec{\Omega},\nu){\rm d}\vec{\Omega}/(4 \pi)$.
Here, $[\bar{\mathcal R}^*]_Q^{KK'}$ denotes the redistribution functions in the comoving frame for the unmagnetic setting.
Third, we only consider the component $K=Q=0$ in the following.
Observing that $\mathcal{T}^0_{0,1} = 1$, we thus rewrite the operator $\Sigma_i^*$ as
\begin{equation*}
\Sigma_i^*=\Upsilon_i^{\bar\nu\nu}\bar\Sigma_i^*\Upsilon_i^{\nu\bar\nu}=\Upsilon_i^{\bar\nu\nu}\bar\Phi_i\Phi^\Omega\Upsilon_i^{\nu\bar\nu},
\end{equation*}
where $\Upsilon_i^{\nu\bar\nu}\in\mathbb R^{N_\Omega N_\nu\times N_\Omega N_\nu}$ and $\Upsilon_i^{\bar\nu\nu}\in\mathbb R^{N_\Omega N_\nu\times N_\nu}$ denote the local interpolations from the observer's grid $\nu$ to the comoving grid $\bar\nu$ and vice versa, respectively, 
$\bar\Phi_i \in \mathbb R^{N_\nu \times N_\nu}$ encodes the linear operator $\mathcal{L}_i[f(\rr_i,\nu)] \! = \! k(\rr_i) \!\int\![\mathcal{\bar R}^*]^{00}_0(\rr,\nu',\nu)f(\rr_i,\nu'){\rm d} \nu'$
and $\Phi^\Omega\in\mathbb R^{N_\nu\times N_\Omega N_\nu}$ represents the angular quadrature (i.e. $\Phi^\Omega \bm{I}_i$ is the local mean intensity in the observer's frame, $J^0_0$, and the linear operator $\bar\Sigma^*=\bar\Phi_i\Phi^\Omega\in\mathbb R^{N_\nu\times N_\Omega N_\nu}$ encodes the local scattering intensity emissivity in the comoving frame for the unpolarised and unmagnetic setting).
Fourth, by multiplying both sides of Eq.~\eqref{eq:Delta_I} by $\Phi^\Omega \Upsilon_i^{\nu\bar\nu}$, we obtain
\begin{equation}\label{eq:A_Block_J}
(I\hspace{-0.1em}d^{N_\nu}-A_i)\Phi^\Omega \Upsilon_i^{\nu\bar\nu}(\bm{I}_i^{n+1}-\bm{I}_i^{n})=\Phi^\Omega \Upsilon_i^{\nu\bar\nu}\bm{R}_i^n,
\end{equation}
with $I\hspace{-0.1em}d^{N_\nu}\in\mathbb R^{N_\nu\times N_\nu}$ the identity matrix and where we introduced the invertible square matrix $A_i=\Phi^\Omega \Upsilon_i^{\nu\bar\nu}\Lambda_{ii}\Upsilon_i^{\bar\nu\nu}\bar\Phi_i\in\mathbb R^{N_\nu\times N_\nu}$.
We finally use Eq.~\eqref{eq:A_Block_J} to write
\begin{align*}
\Sigma_i^*(\bm{I}_i^{n+1}-\bm{I}_i^n)=&\,\Upsilon_i^{\bar\nu\nu}\bar\Phi_i\Phi^\Omega\Upsilon_i^{\nu\bar\nu}(\bm{I}_i^{n+1}-\bm{I}_i^n)\\
=&\,\Upsilon_i^{\bar\nu\nu}\bar\Phi_i(I\hspace{-0.1em}d^{N_\nu}-A_i)^{-1}\Phi^\Omega \Upsilon_i^{\nu\bar\nu}\bm{R}_i^n,
\end{align*}
and replace it in Eq.~\eqref{eq:Block_Jacobi} to obtain
\begin{equation*}
\bm{I}_i^{n+1}-\bm{I}_i^{n}=\bm{R}_i^n+\Lambda_{ii}\Upsilon_i^{\nu\bar\nu}\bar\Phi_i(I\hspace{-0.1em}d^{N_\nu}-A_i)^{-1}\Phi^\Omega\Upsilon_i^{\nu\bar\nu}\bm{R}_i^n.
\end{equation*}
From this expression, we can identify the local blocks of $\hat{P}^{-1}$ as
\begin{equation*}
\hat P_i^{-1}=I\hspace{-0.1em}d^{N_\Omega N_\nu}+\Lambda_{ii}\Upsilon_i^{\nu\bar\nu}\bar\Phi_i(I\hspace{-0.1em}d^{N_\nu}-A_i)^{-1}\Phi^\Omega\Upsilon_i^{\nu\bar\nu},
\end{equation*}
where the matrices $(I\hspace{-0.1em}d^{N_\nu}-A_i)^{-1}\in\mathbb R^{N_\nu\times N_\nu}$ can be assembled a priori, before starting the iterative method.
We note that, for the block-diagonal preconditioner, the continuum contributions to $\Sigma_i^*$ were neglected for computational efficiency.

\end{document}